\documentclass[preprint,12pt]{elsarticle}

\usepackage{graphicx}
\usepackage{amssymb}

\usepackage{lineno}
\usepackage{amsmath}
\usepackage{makecell}

\journal{Journal Name}

\begin{document}

\begin{frontmatter}


\title{Renewable Generation Data for European Energy System Analysis}



\author{Alexander~Kies$^{1}$, Bruno~U.~Schyska$^{2}$, Mariia~Bilousova$^{1,3}$, Omar~El~Sayed$^{1,3}$, Jakub~Jurasz$^{4,5,6}$, Horst~Stoecker$^{1,3,7}$}

\address{%
$^{1}$Frankfurt Institute for Advanced Studies, Goethe-University Frankfurt, Ruth-Moufang-Str. 1, 60438 Frankfurt, Germany\\
$^{2}$German Aerospace Center (DLR), Institute of Networked Energy Systems, Carl-von-Ossietzky-Straße 15, 26129 Oldenburg, Germany\\
$^{3}$Goethe-University, Frankfurt, Germany\\
$^{4}$Faculty of Environmental Engineering, Wroclaw University of Science and
Technology, 50-370 Wroclaw, Poland\\
$^{5}$Faculty of Management, AGH University, 30-059 Cracow, Poland\\
$^{6}$School of Business, Society and Engineering, Mälardalens Högskola, 72113
Västerås, Sweden\\
$^{7}$GSI Helmholtzzentrum für Schwerionenforschung, Planckstraße 1, 64291 Darmstadt, Germany\\}

\begin{abstract}
In the process of decarbonization, the global energy mix is shifting from fossil fuels to renewables. To study decarbonization pathways, large-scale energy system models are utilized. These models require accurate data on renewable generation to develop their full potential.
 Using different data can lead to conflicting results and policy advice. In this work, we compare several datasets that are commonly used to study the transition towards highly renewable European power system. We find significant differences between these datasets and cost-difference of about 10\% result in the different energy mix. We conclude that much more attention must be paid to the large uncertainties of the input data.
\end{abstract}

\begin{keyword}
Energy System Analysis \sep Renewable Generation Data \sep Energy Meteorology \sep MERRA-2 \sep ERA5, renewables.ninja, EMHires


\end{keyword}

\end{frontmatter}


\section{Introduction}
\label{S:1}
Sustainable energy sources are a major solution for the imminent thread of climate change \cite{chu2012opportunities,change2014mitigation}.
Increasing the number of installations of renewable generation capacities and the electrification of energy sectors like heating and transportation fosters decarbonization. As part of the Paris Agreement, many countries around the world commited to reducing their greenhouse gas emissions. In its turn, the EU set an aim to reduce emissions by at least 50\% by 2030, as compared to 1990 levels \cite{greendeal}. 
In this context, the sensitivity of energy systems to weather and climate rises \cite{bloomfield2020importance,staffell2018increasing}. Energy system models play an important role in understanding and investigating energy systems of various scales and scopes. For the renewable future, the use of adequate meteorological data in the field of energy system analysis and modelling is essential.

Common large-scale energy models for Europe cover the EU \cite{muller2018comprehensive} and aggregate quantities such as generation from renewable sources to country levels. 
There are multiple datasets that provide data on hourly generation potentials from renewable sources such as wind and solar PV. They are based on reanalysis data, which is an assimilation of historical measurements and numerical models into a consistent estimate of a state of the atmosphere. These datasets provide relevant variables such as wind speed, irradiation or temperature used to compute potential generation from renewables. 

A comparison of two of these datasets was performed by Moraes et al. \cite{moraes2018comparison}. They found that datasets diverge from each other even if they are based on the same meteorological source. This is likely based on differences in technological assumptions that are made to convert meteorological data to generation. These include, for instance, assumptions on the wind turbines used, which can take in historical data or projections into the future such as increasing hub heights \cite{lantz2019increasing}, as well as their placement. 

In this paper, we compare seven datasets which provide time-resolved aggregated generation data on the country level. Only five of them cover the same time period (2003-2012), therefore we focus on them in particular.
Section 2 discusses these datasets and shows the major steps in converting meteorological data to energy system model input. Section 3 analyses the datasets with respect to different means such as annual capacity factors, ramp rates and optimised mixes. Section 4 discusses the results and implications and Section 5 concludes this paper.

This paper contains novelties relevant for the research community. It provides the broadest comparison of renewable generation datasets to date, introduces energy system-related measures to compare them and derives important conclusions for research on the energy transition.

\section{Data}
Data on renewable generation potentials is essential for the modelling and analysis of energy systems with significant shares of generation from renewable sources. Among the most used datasets in the study of weather and climate are reanalyses. A meteorological reanalysis is a method to create long-term weather data using numerical weather prediction models and assimilating historical data. Reanalyses are used to study climate variability \cite{kravtsov2014two} as well as are commonly employed to study energy systems
\cite{jurasz2020review}. Wind speeds for the datasets investigated in this work are mostly based on two reanalyses: ERA5 and MERRA-2.
MERRA \cite{rienecker2011merra} is provided by NASA. The official data production was
launched in 2008 with the use of the up-to-date GEOS-5 (Goddard Earth Observing System Data Assimilation
System Version 5) produced at NASA GMAO (Global Modeling and Assimilation Office).
ERA5 \cite{hersbach2020era5} is the newest global reanalysis by the European Centre for Medium-Range Weather Forecasts (ECMWF). Wind speed data are provided for heights of 10 m as well as 100 m.
The RE-Europe is the only considered dataset that uses ECMWF forecasts (RE-Europe) and the regional COSMO-REA6 reanalysis for the modelling of wind power \cite{bollmeyer2015towards}.
For solar data, CM SAF SARAH is commonly used alongside the reanalyses  \cite{schulz2009operational}. It is a satellite-based climate data record of irradiance data and other variables.

MERRA-2 and ERA5 were compared by various researchers. Olauson \cite{olauson2018era5} compared ERA5 and MERRA-2 for the modelling of wind power on a country-level as well as for individual turbines. His findings indicate that ERA5 performs considerably better than MERRA-2 on both levels.
Camargo et al. \cite{camargo2020simulation} used ERA5 data to model multi-annual time series of solar PV generation. They performed a validation with hourly data of PV plants in Chile and found a slightly superior performance of ERA5 as compared to modelling based on MERRA-2.
Gruber et al. \cite{gruber2020towards} compared MERRA-2 and ERA5 for wind power simulation bias-corrected with the global wind atlas for the US, Brazil, New Zealand and South Africa and found ERA5 to outperform MERRA-2.
Urraca et al. \cite{urraca2018evaluation} evaluated global horizontal irradiance estimates from ERA5 and COSMO-REA6, a regional reanalysis from the German Meteorological Service (DWD), and concluded that both reanalyses reduce the quality gap between reanalysis and satellite data.
Jourdier \cite{jourdier2020evaluation} investigated ERA5, MERRA-2 and other datasets to examine wind power production in France, and found that ERA5 is skilled, however, it underestimates wind speeds, especially in mountainous areas.
Piasecki et al. \cite{piasecki2019measurements} compared ERA5 data with measurements in several locations in Poland and concluded that  they are in good agreement for solar PV, with hourly correlation coefficients above 0.9, while wind comparison showed a large variability with differences in capacity factors of up to 15 percentage points.

\begin{table}[!h]
\begin{center}
\makebox[\textwidth][c]{
\begin{tabular}{lllllll}
Dataset          & years & wind & PV & \makecell{capacity\\layout} & \makecell{temporal\\ resolution} & ref. \\\hline
renewables.ninja &  1980-2019             &  MERRA         &    Sarah     &   yes                   & 1h          &        \cite{pfenninger2016renewables}   \\
EMHires          &  1986-2015             &  MERRA         &    Sarah     &        yes                        &     1h  &  \cite{gonzalez2016emhires}  \\
Restore          &  2003-2012             &  MERRA         &    Sarah    &  no                              &    1h  &    \cite{kies2016restore}   \\
UReading-E          &  1979-2018             &   ERA5        &  ERA5  &      yes                          &       3h  &  \cite{bloomfield2020era5}  \\
UReading-M          &  1979-2018             &   MERRA        & MERRA       &      yes                          &       1h  &  \cite{bloomfield2020merra2}  \\
PyPSA-Eur        &  2013             &     ERA5      &     Sarah    &             yes                   &       1h  &  \cite{horsch2018pypsa}  \\
RE-Europe        &  2012-2014             &    diff.       &   diff.      &      no                        &   1h  &      \cite{jensen2017re} 
\end{tabular}
}
\end{center}
\caption{\label{table:datasets}Datasets providing data on renewable generation that were analysed and compared in this work. Abbreviations used in this work are Ninja, URead-E/M and RE-Eur-E/C.}
\end{table}

\begin{figure}
    \centering
    \includegraphics[width=0.32\textwidth]{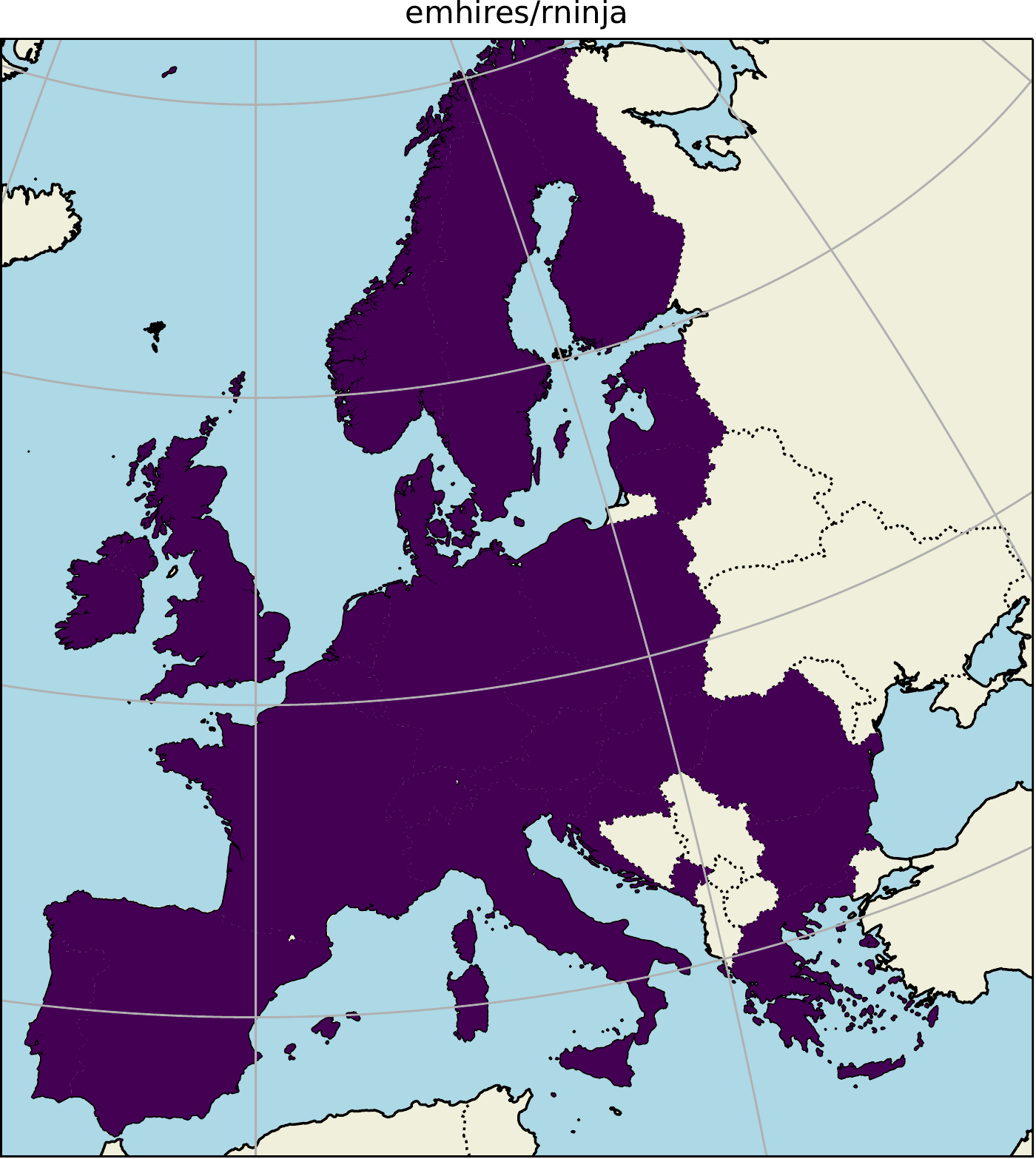}
    \includegraphics[width=0.32\textwidth]{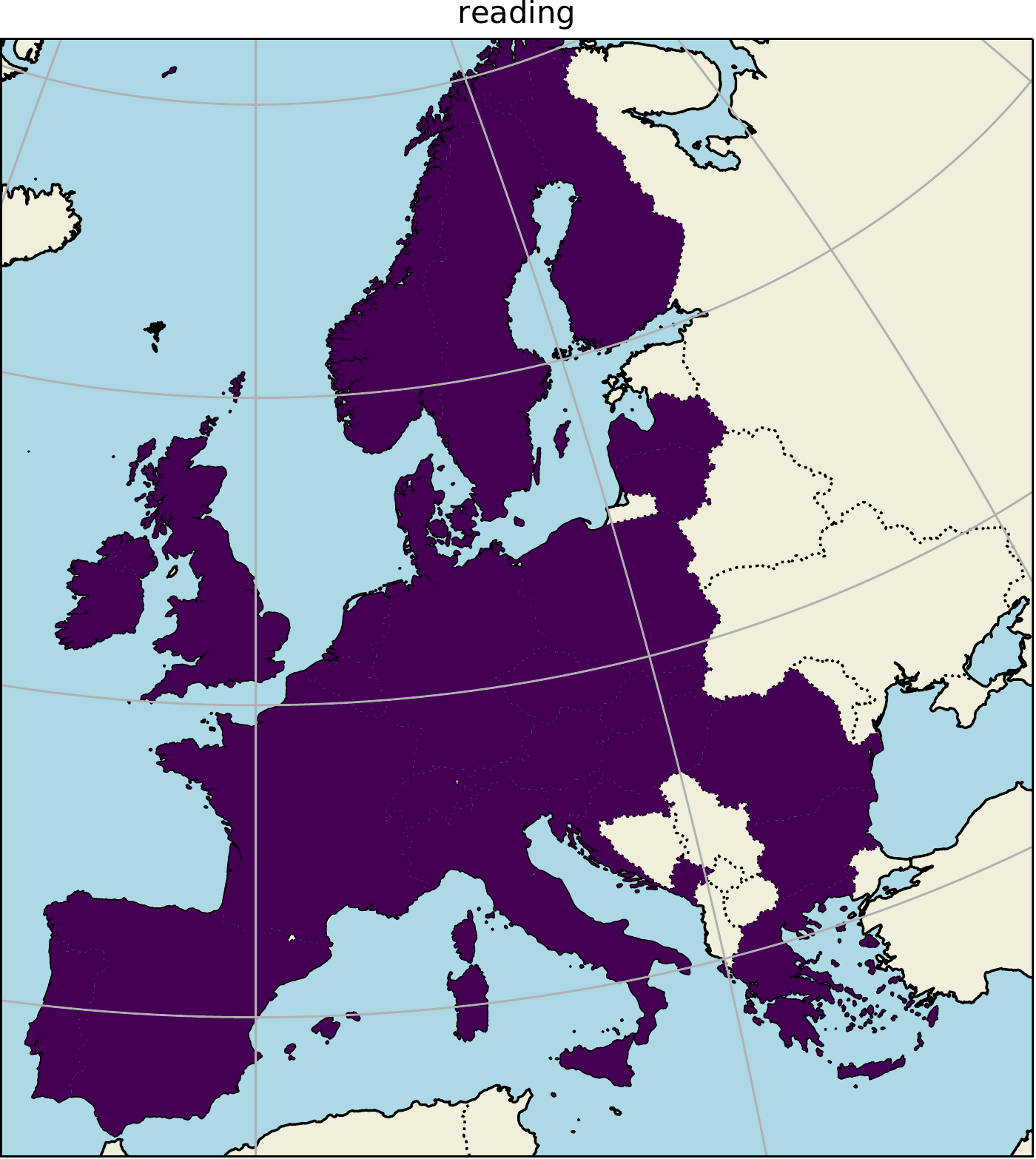}
    \includegraphics[width=0.32\textwidth]{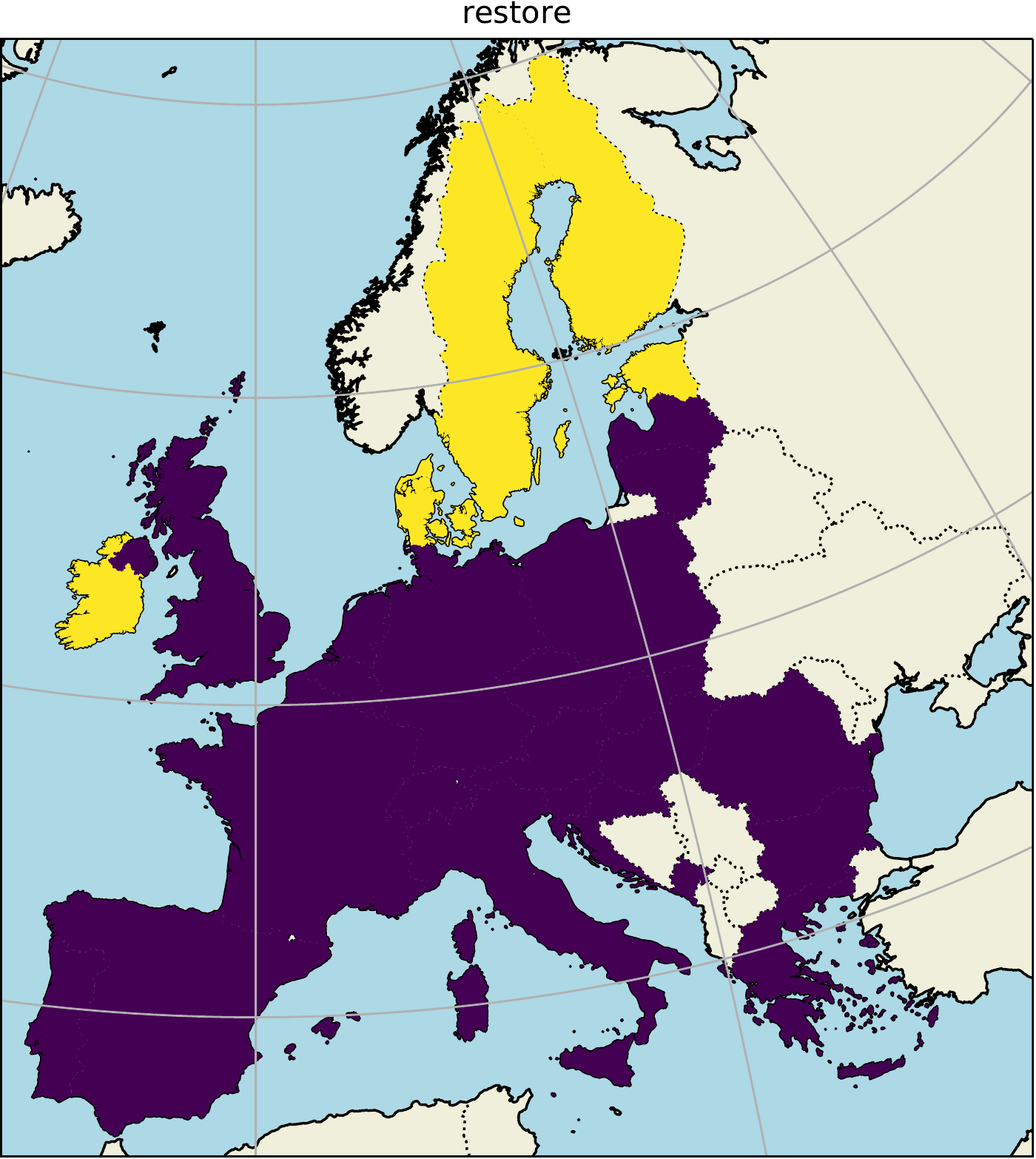}
    
    \caption{Country coverage of the datasets. Purple-colored countries include both solar PV and wind time series, yellow-colored ones include only wind. Both UReading datasets exclude Estonia. The Restore dataset excludes Norway and for countries Ireland, Denmark, Finland and Estonia provides only wind time series. }
    \label{fig:country_coverage}
\end{figure}
Table \ref{table:datasets} shows key features of each dataset.
Figure \ref{fig:country_coverage} shows the countries covered.
To use a sufficiently long time period for every dataset, we set the time frame of the following analyses to the period 2003-2012 and compare 5 datasets covering it. 

\begin{figure}[!htp]
    \centering
    \includegraphics[width=0.7\textwidth]{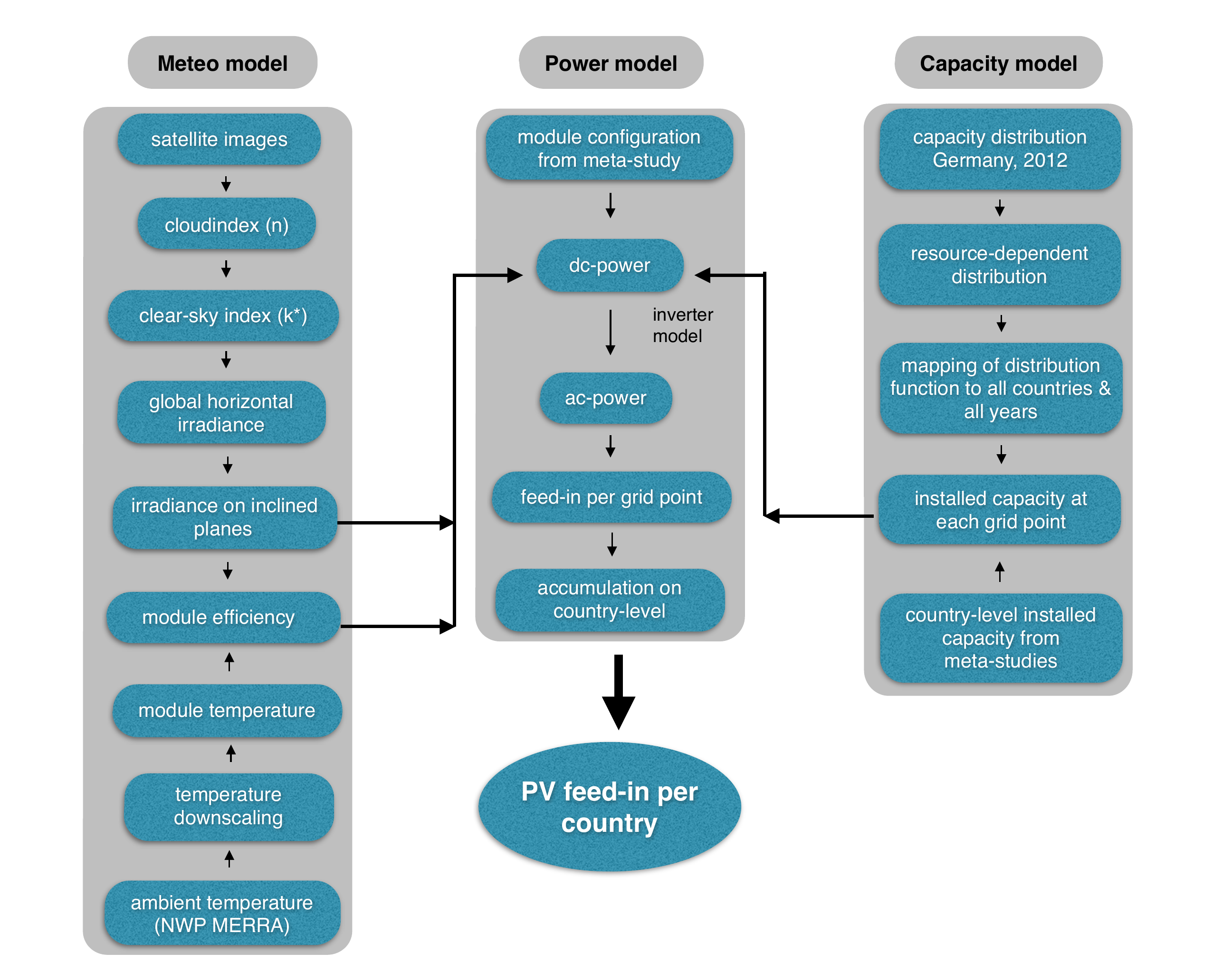}\\
     \includegraphics[width=0.7\textwidth]{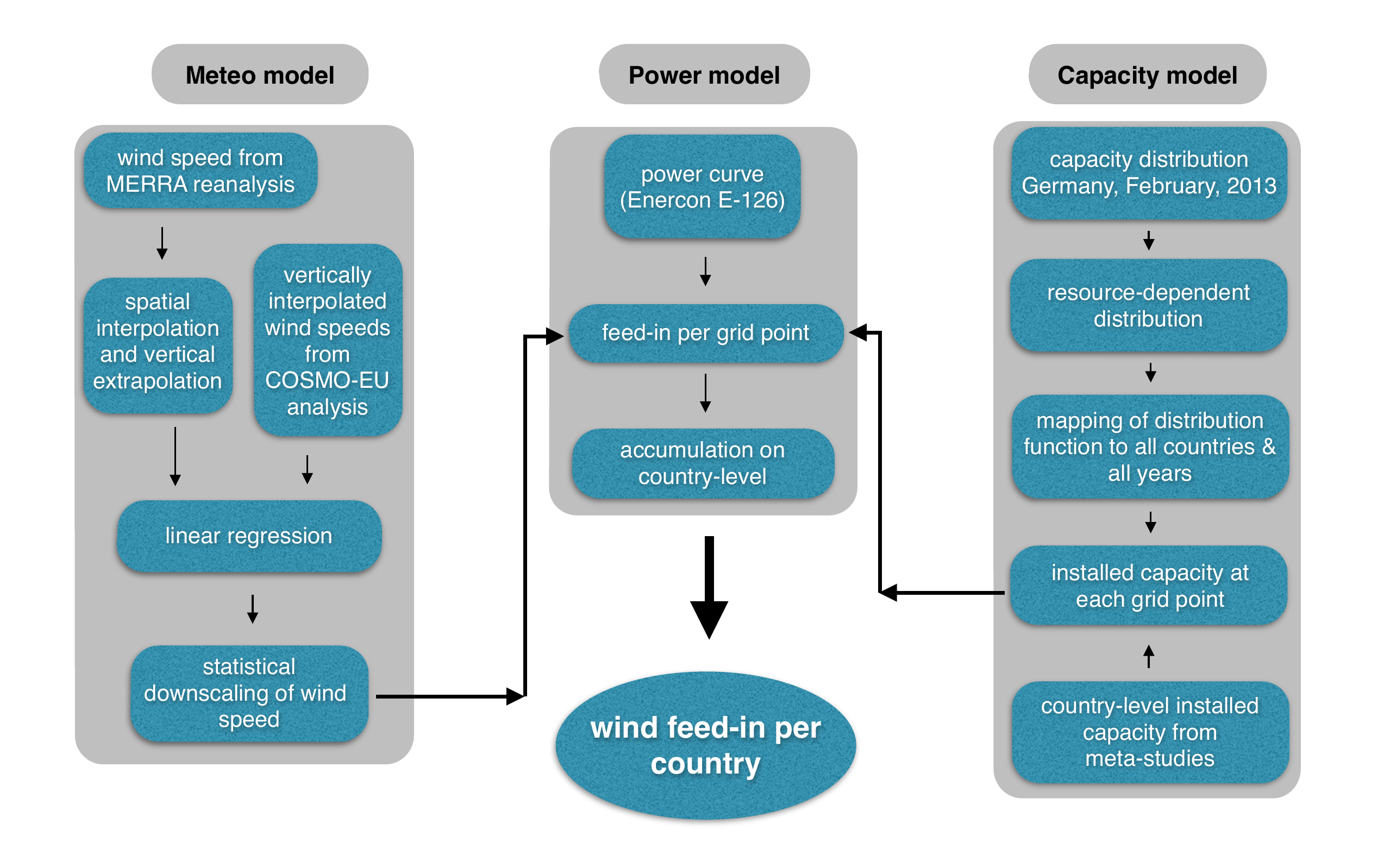}
    \caption{Schematic  flowchart for PV/wind generation modelling for the RESTORE dataset. Figure from Kies et al. \cite{kies2016restore}.} 
    \label{fig:my_flowchart}
\end{figure}

For the sake of readability, we only briefly summarize the major steps of modelling renewable energy generation for the datasets investigated in this section, except for the Restore dataset, where we describe them in more detail. In general, similar steps are taken for all datasets  (e.g. Fig.~ \ref{fig:my_flowchart}). A meteo model converts meteorological data to the needed input for a power model, which yields space- and time-dependent power per installed capacity. The output of this model is aggregated by a capacity model, which represents how much capacity is installed at what location. The aggregated sum of this model yields the generation time series for the renewable generation.

\subsection{Restore}
The Restore dataset \cite{kies2016restore} was produced in the context of the RESTORE 2050 project. Besides wind and solar PV generation time series, it also contains time series of other generation technologies, such as CSP, wave and hydro inflow.
To calculate wind power per grid cell in the Restore dataset, wind speeds are interpolated to the grid of the COSMO-EU model \cite{schulz2014kurze}, and extrapolated to the turbine hub height using the wind speed log profile

    \begin{align}
      \frac{s(z_1)}{s(z_2)} &= \frac{\log(z_1/z_0)}{\log(z_2/z_0)},
    \end{align}
    where $z_1,z_2$ are the heights at which wind speeds are given and desired, respectively and $z_0$ is the surface roughness length provided by the dataset.
    Wind speeds are corrected using a linear regression according to COSMO-EU wind speeds. 
The power curve of an Enercon E-126 turbine is then used to convert wind speeds to wind power at 100m hub height. Additionally, the dataset provides capacity factors generated with the same power curve for the hub heights of 140m and 180m. In the following comparison, we use  the dataset based on 100m wind speeds.
Any wind turbine power curve mapping wind speed $s$ to power output $g$ is approximately given by \cite{lydia2014comprehensive,taslimi2016development}

\begin{align}
    g = 
    \begin{cases}
0 \text{ for } s < s_0 \vee s > s_\text{max}\\
\propto \left(s-s_0\right)^3 \text{ for } s_0 < s < s_\text{nom}\\
= G \text{ for } s_\text{nom} < s < s_\text{max}.
\end{cases}
\end{align}

 For low wind speeds, torque caused by the wind on blades of a turbine is too low to generate angular momentum. Above a certain speed, referred to as cut-in speed $s_0$, turbines start to rotate and produce power. The power curve now resembles the characteristic $v^3$-dependency of the kinetic energy of wind passing through a certain area. At the rated wind speed $s_\text{nom}$, the rated power of the turbine is reached. This power is kept constant by the turbine beyond this speed value, commonly by the adjustment of blade angles. If speeds increase further, they can reach a critical level, referred to as cut-out speed $s_\text{max}$, at which the rotor blades are turned out of the wind to prevent a structural damage to the turbine.

For PV Power, global horizontal irradiance is converted to irradiation on inclined surfaces  based on the Klucher model \cite{klucher1979evaluation}. For tilt angles, optimized values per country are used.

The efficiency of the PV modules in dependency of incoming irradiation $I$ and temperature $T$ is modelled via a parametric model for a standard temperature

  \begin{align}
      \eta(I, 25^\circ C) = a1 + a2 \times I + a3 \ln I,
      \label{eqn:eta_mpp_25}
  \end{align} 
  where $a1,a2,a3$ are device-specific parameters and the temperature dependency of the efficiency is given via
  
   \begin{align}
      \eta(I, T) = \eta(I, 25^\circ C)(1 - 0.004 \Delta T).
      \label{eqn:eta_mpp_tm}
   \end{align}

PV power output is then directly computable via

  \begin{equation}
    \begin{split}
      g_t & = \eta_t I_t A.
     \label{eqn:Itilt_power}
    \end{split}
  \end{equation}
  
  Unlike the other datasets investigated in this work, the Restore dataset does not take the locations of existing generation capacities into account. Instead, country-wise capacities are distributed proportionally to the underlying resource \cite{kies2016restore}.
  
\subsection{UReading}
The UReading datasets \cite{bloomfield2020era5,bloomfield2020merra2} were produced by researchers from the University of Reading, UK. They contain hourly values of aggregated power generation from wind and solar based on a representative distribution of wind and solar farms as well as ERA5 (Reading-E) or MERRA-2 (Reading-M) reanalyses data for 28 European countries. In addition, a daily time series of electricity demand is provided. It is used in this paper to optimise generation mixes.

\subsection{Renewables.ninja}
Renewables.ninja \cite{pfenninger2016renewables} is a web tool that provides potential generation of wind and solar PV for single locations or countries globally. Pfenninger and Staffell  \cite{pfenninger2016long,staffell2016using} also found that significant correction factors are necessary to model renewable feed-in from reanalyses in Europe. To model PV power, they use the model by Huld et al. \cite{huld2010mapping} with irradiation based on SARAH for different azimuth/tilt combinations.
For wind power, they use a virtual wind farm model \cite{staffell2014does} and bias-correct it.

\subsection{EMHires}
European   Meteorological   derived   HIgh resolution   RES generation time series for present and future scenarios (EMHires) is a dataset produced by the joint research centre (JRC) with an aim to  allow  users  to  assess  the  impact  of  meteorological and climate variability on renewable generation in Europe \cite{gonzalez2016emhires,gonzalez2017simulating,gonzalez2017emhires}.
For wind power, this dataset combines MERRA-2 reanalysis data with wind farms data from thewindpower.net \cite{windpower}. Wind power time series are further normalised to the reported ENTSO-E annual production statistics. Wind speeds are statistically downscaled and interpolated to the desired hub height using a wind profile power law. Wind speed data are then converted to wind power using a specific power curve assigned  to each  wind  farm  considering  the  characteristics  of  the wind  farm, such as manufacturer.
For PV generation, the PVGIS model \cite{huld2015estimating,huld2012new,vsuri2007potential} is used to provide generation in dependency of irradiation and solar module parameters. Together with assumptions such as inclinations, PVGIS output is then aggregated to country levels.

\subsection{PyPSA-Eur}
PyPSA-EUR is a dataset for European generation and transmission expansion planning studies from freely available data \cite{horsch2018pypsa}. Besides time series of renewable generation, it contains additional data to model a renewable European energy system, such as alternating/direct current transmission lines, substations, data on conventional generators and demand as well as renewable capacity installation potentials.
For renewable generation data, it uses data from the ERA5 reanalysis with a logarithmic wind power profile to extrapolate wind speeds to the desired hub height for wind power modelling. For PV conversion, it also uses the PVGIS model \cite{huld2012new}.

\subsection{RE-Europe}
RE-Europe \cite{jensen2017re} is a dataset produced at the Technical University of Denmark for the modelling of a highly renewable European power system. Analogously to PyPSA-EUR, alongside renewable generation data it contains other data, for instance data on transmission lines. To provide meteorological variables for both, wind and solar PV, meteorological data from the ECMWF forecasts as well as the COSMO-REA6 reanalysis are used.
For the final conversion step, it uses a smoothed power curve of a specific wind turbine (Siemens SWT 107) as well as a specific solar panel (Scheuten P6–54 215 Multisol Integra Gold) to convert meteorological variables to a potential generation.
Generation capacities are assigned to sub-national regions using two heuristic capacity layouts.






\section{Results}
In this section, we study different properties of the provided time series of generation per unit $g_{n,s,t}^i$. Here $i$ denotes the dataset, $n$ is the country, $s$ is the technology and $t$ is the timestep.
\subsection{Annual Capacity Factors}

The capacity factor of a generator is the ratio of the potential energy output over a given period of time to the maximum possible energy output as given by the rated capacity over that period \cite{abed1997capacity},

\begin{align}
    \text{cf} &= \sum_t\frac{g^+_{n,s,t}}{G_{n,s}}.
\end{align}

For renewable generators the capacity factor heavily depends on the resource availability as well as on the technical parameters. Occasionally, it is referred to as full load hours. Besides reductions due to resource inavailability, one can, for instance, factor net congestions into effective capacity factors \cite{kies2016curtailment}.

\begin{figure}
    \centering
    \includegraphics[width=\textwidth]{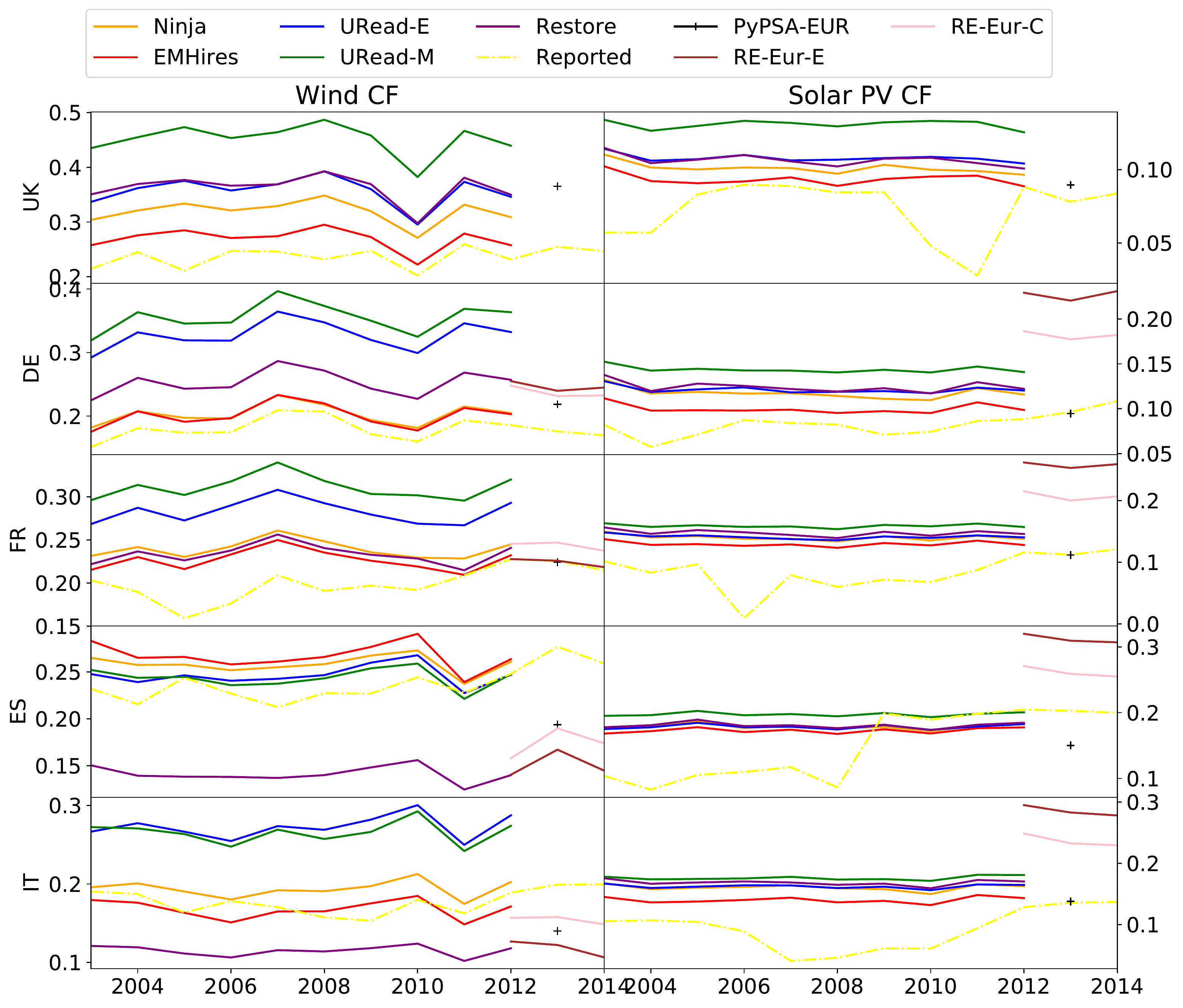}\\
    \caption{Annual capacity factors for wind and solar PV for the chosen countries. RE-Europe does not cover the UK. Solar PV capacity factors are the highest for the RE-Europe dataset. Year-to-year changes are well caught by all datasets for wind, implying the difference resulting from the conversion process from meteorological data to generation. For solar PV, capacities in some countries such as the UK and Italy were very small (well below 1 GW) at the beginning of the investigated period. Strong changes in year-to-year reported capacity factors are potentially just caused by rounding or reporting errors. Also taking the large year-to-year changes in reported capacity factors into account, it seems questionable, how valueable these values are in many cases for a comparison with modelled data. }
    \label{fig:cfs}
\end{figure}
\begin{figure}
    \centering
    \includegraphics[width=\textwidth]{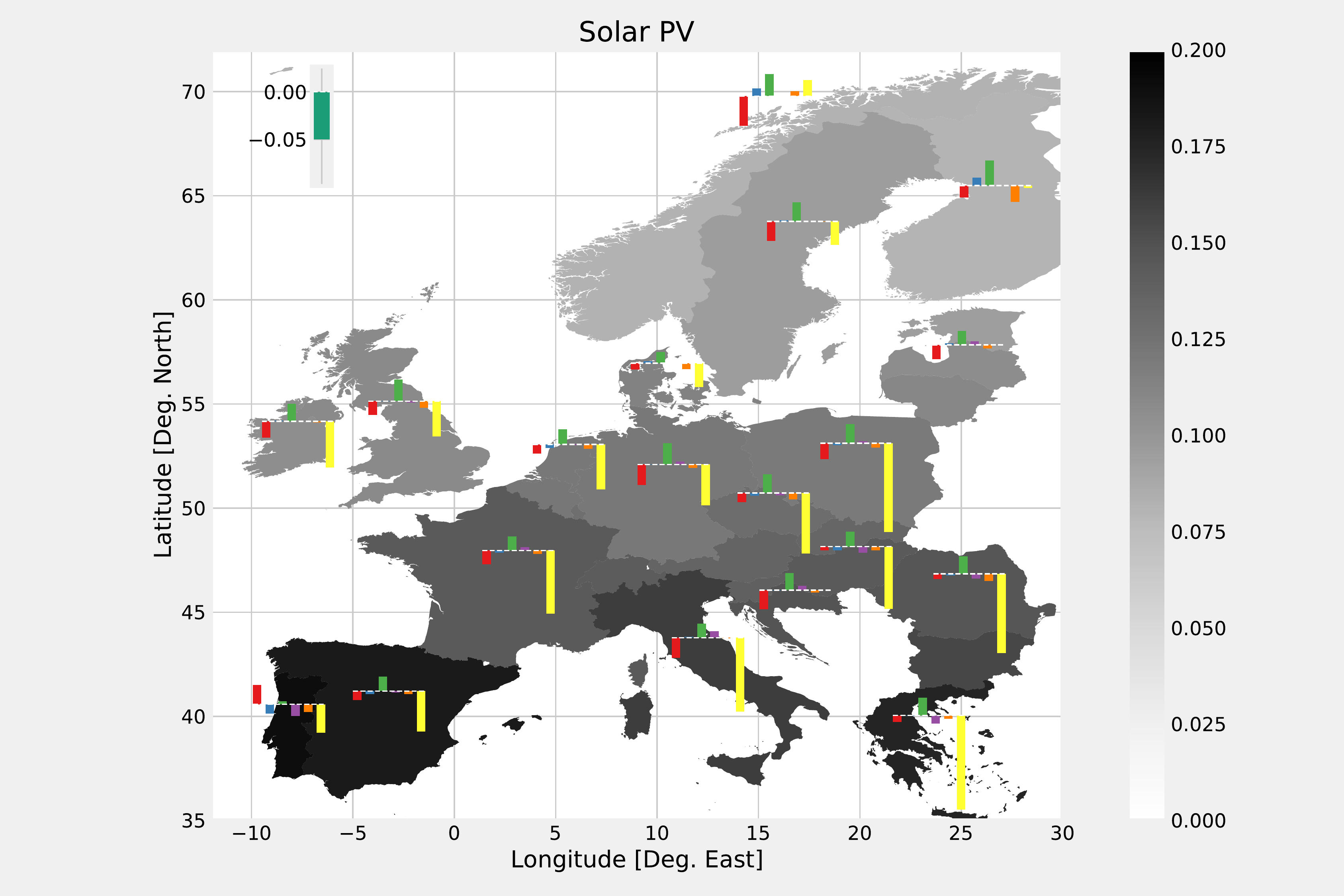}\\
     \includegraphics[width=\textwidth]{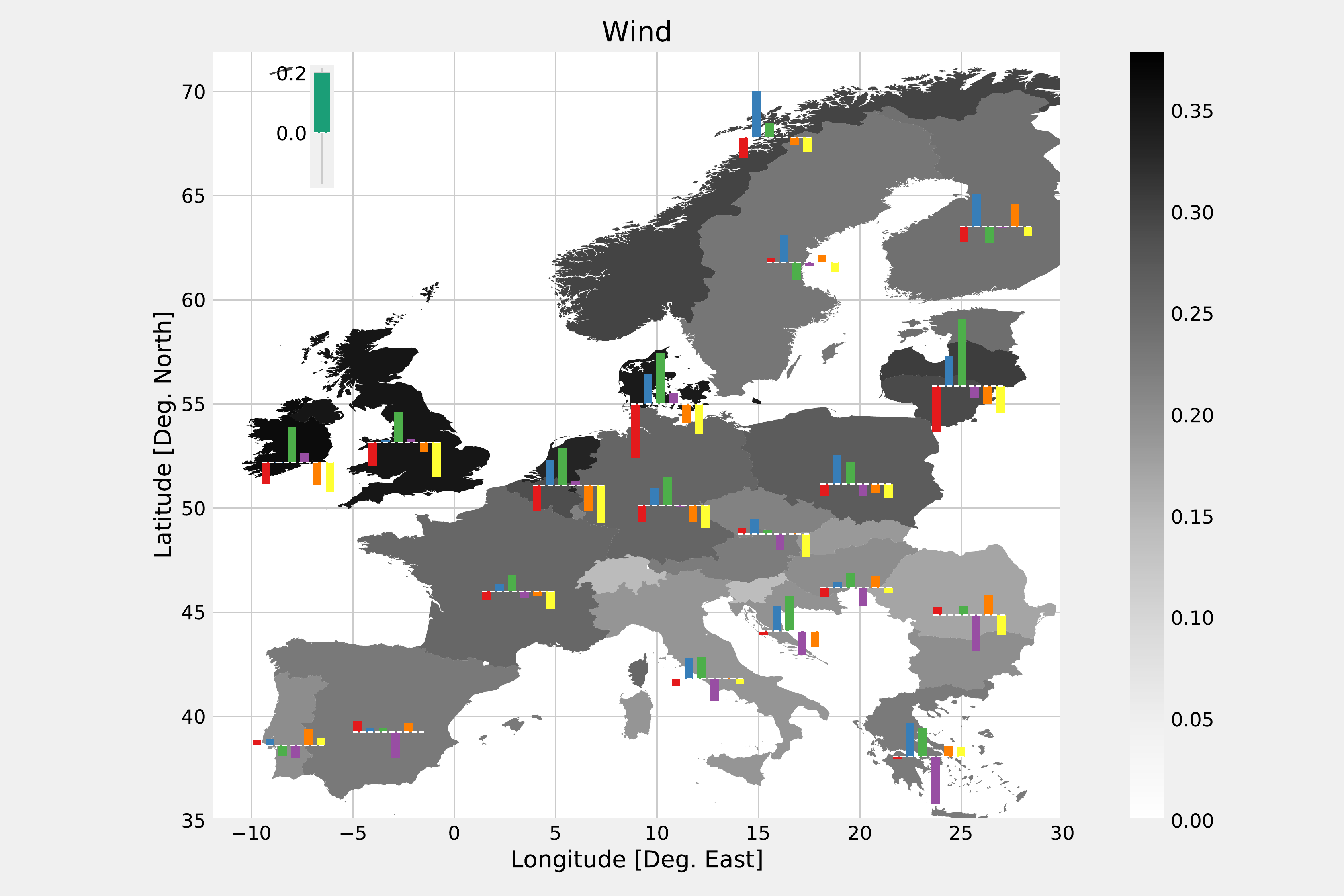}
    \caption{Average capacity factors for wind and solar PV in the years 2003-2012. The greyscale bar indicates the average capacity factor. The colorbar plots per country show the average deviation in percentage points from the mean of the ensembles (Restore, EMHires, URead-E, URead-M and Ninja). Colors are: EMHires (red), URead-E (blue), URead-M (green), Restore (purple), Ninja (orange) and reported (yellow). For solar PV, trends seem to be consistent over all for Europe. For wind, there is more variation. For instance, Restore produces comparably small capacity factors in South-East Europe and Norway, while for other North Sea countries Restore capacity factors are above average.}
    \label{fig:cfs_maps}
\end{figure}
Figure \ref{fig:cfs} shows annual capacity factors of the different datasets. 
Annual reported capacity factors were calculated by dividing reported generation by reported installed capacities obtained from IRENA \cite{irena}. 
For wind, overall capacity factors differ significantly between datasets, while relative year-to-year changes are quite similar. Reading-M in general shows the highest capacity factor in almost all cases for wind and solar PV, while EMHires shows the lowest. 
Absolute capacity factor differences go up to around 20\% (UK wind), and relative capacity factors for some cases differ by a factor of almost 3 (Italy wind). The UReading-M and UReading-E dataset have different meteorological reanalyses as input but are based on the same assumptions. They show a relatively good agreement for wind and are closest to each other in all cases, except for the UK. For solar PV, UReading-M has relatively low agreement with the other datasets. This potentially supports the observation by Camargo et al. \cite{camargo2020simulation} that modelled PV power based on ERA5 has better agreement with measurements than MERRA-2. For wind, differences when switching between MERRA-2 and ERA5 seem to have a smaller effect than when choosing a different methodological approach, as demonstrated by the proximity of capacity factors for UReading-M and -E.

Another important point is the difference of the modelled values to the reported capacity factors. In some cases, such as German wind, reported values are well captured by Ninja and the EMHires dataset, while results for PV seem to not meet the modelled values, in both absolute terms and year-to-year variations. However, one should keep in mind that not all datasets have the goal to reproduce historical values of capacity factors. Another likely interpretation of the fact that year-to-year changes of reported values differ from the modelled dataset is that modelled values are a subject to meteorological year-to-year changes only, while reported values are also influenced by changing capacity layouts or technological parameters.
The discrepancy between modelled and realized capacity factors with the latter being considerably lower has already led to public debates \cite{boccard2009capacity}. 
There is also a discrepancy between wind capacity factor estimates and realized values based on oversimplifcations \cite{miller2018observation}.
For PyPSA-EUR, the year 2013 and for RE-Europe the years 2012-2014 were considered. Capacity factors in both datasets look unremarkable with the exception of solar PV for RE-Europe; its capacity factors are substantially higher in all considered countries based on both, ECMWF forecasts as well as COSMO-REA6. 

Figure \ref{fig:cfs_maps} shows averaged capacity factors per country for five datasets and reported values. Reported values for solar PV are for most countries are significantly below the modelled data mean, except for Sweden and Finland. U-Read-M reports highest solar PV capacity factors for almost all countries. For solar PV, it can be concluded, that the methodology and the chosen datasets are largely independent of the location.
For wind, the picture looks different. Restore, for instance, has mostly lower capacity factors in Southern Europe, while in North- and Northwestern Europe they are close to or above the ensemble mean. For EMHires, the picture looks the opposite, with capacity factors below average in the North and Northwest and close or above average in the South-East.

\subsection{Correlation}
\begin{figure}
    \centering
    \includegraphics[width=0.49\textwidth]{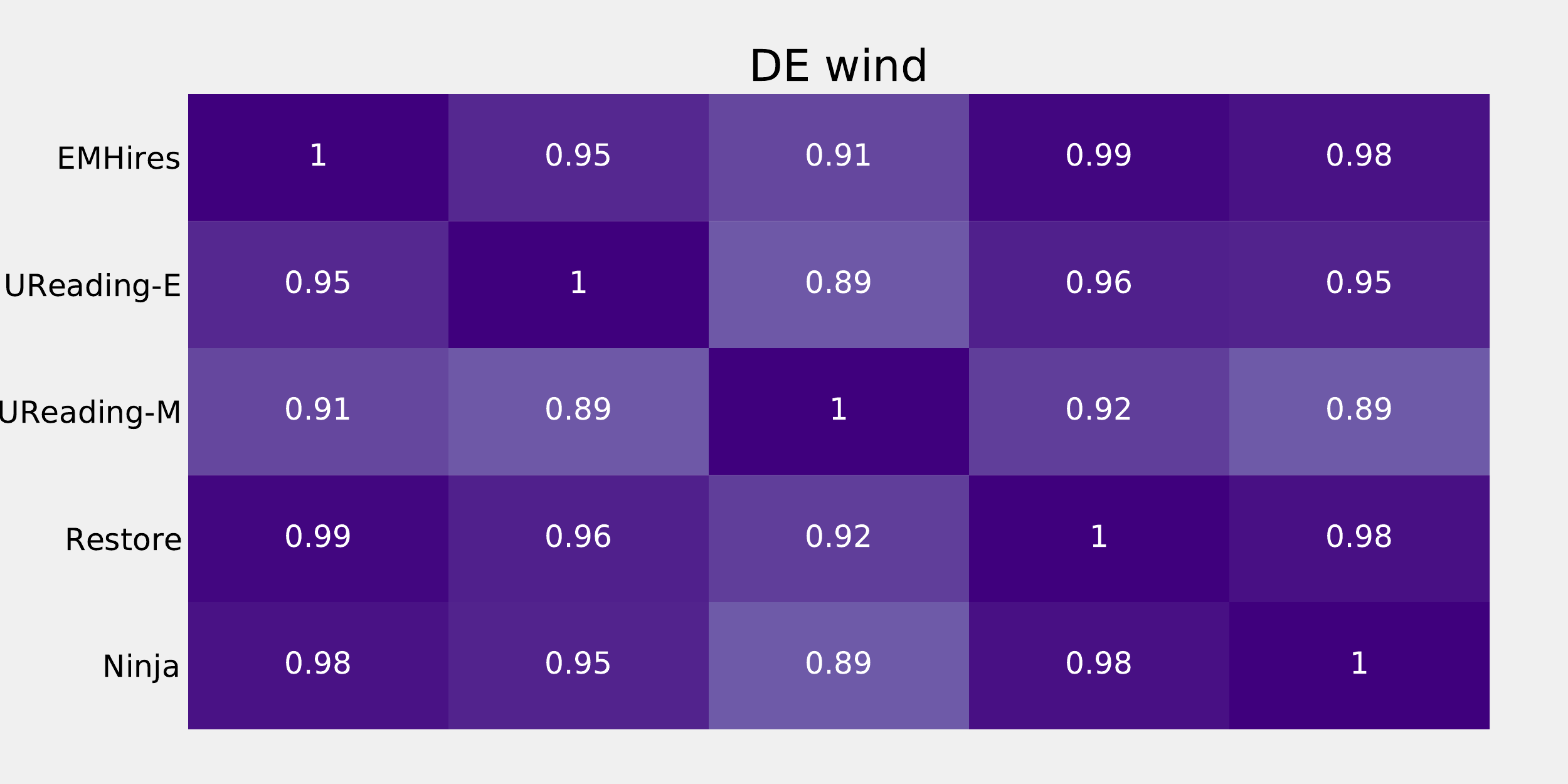}\hspace{-0.2cm}
    \includegraphics[width=0.49\textwidth]{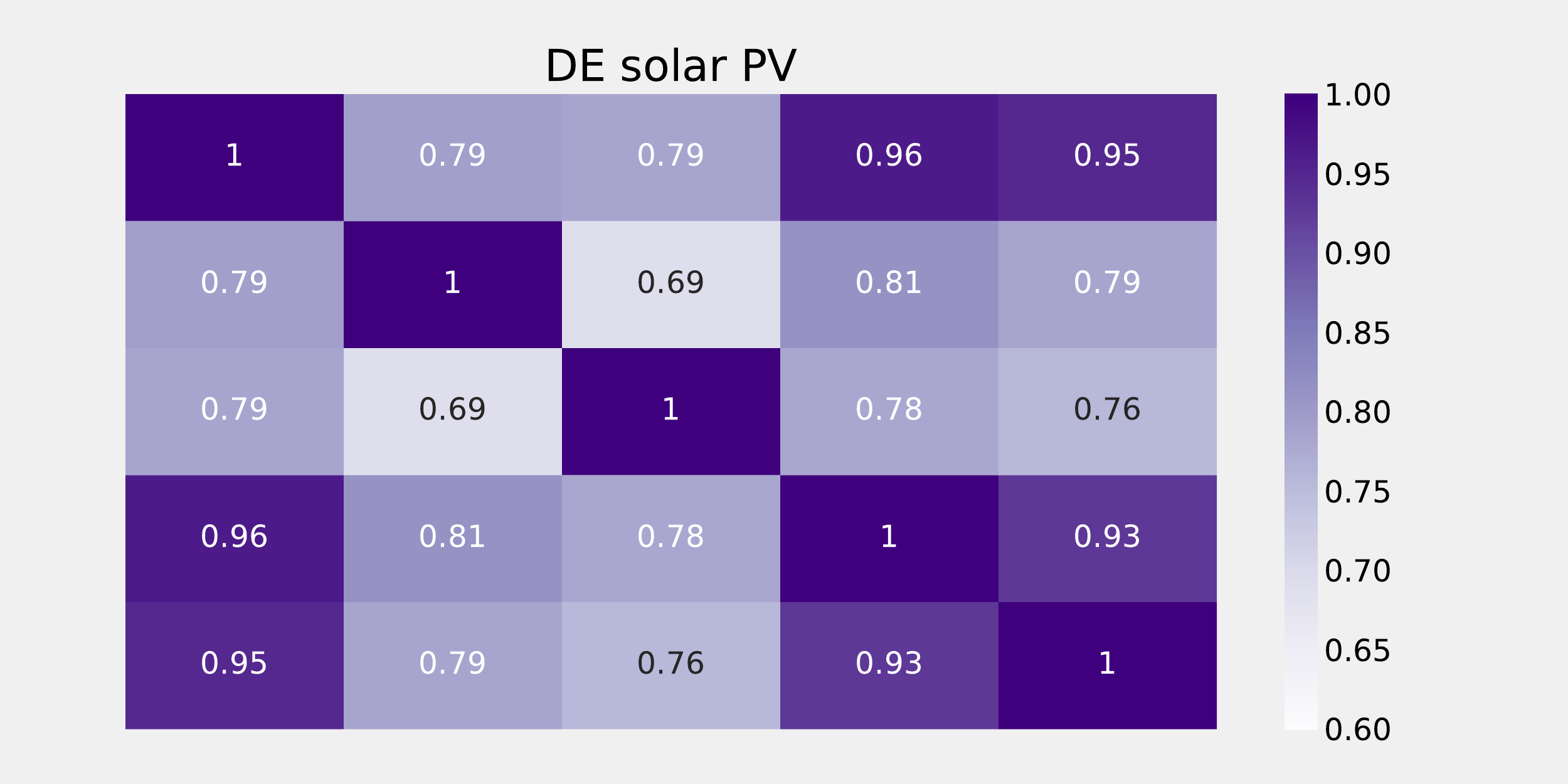}\\\vspace{-0.2cm}
        \includegraphics[width=0.49\textwidth]{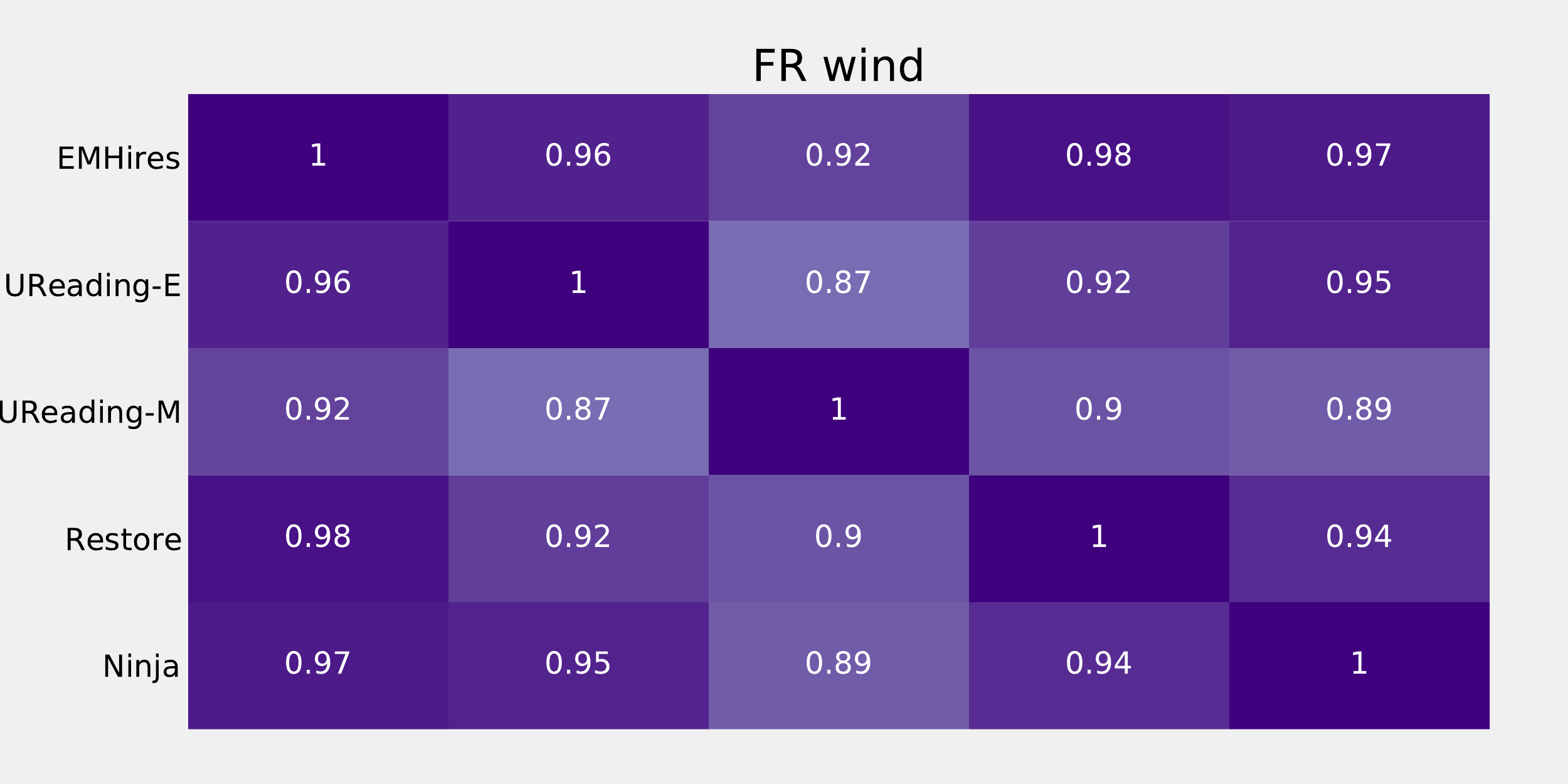}\hspace{-0.2cm}
    \includegraphics[width=0.49\textwidth]{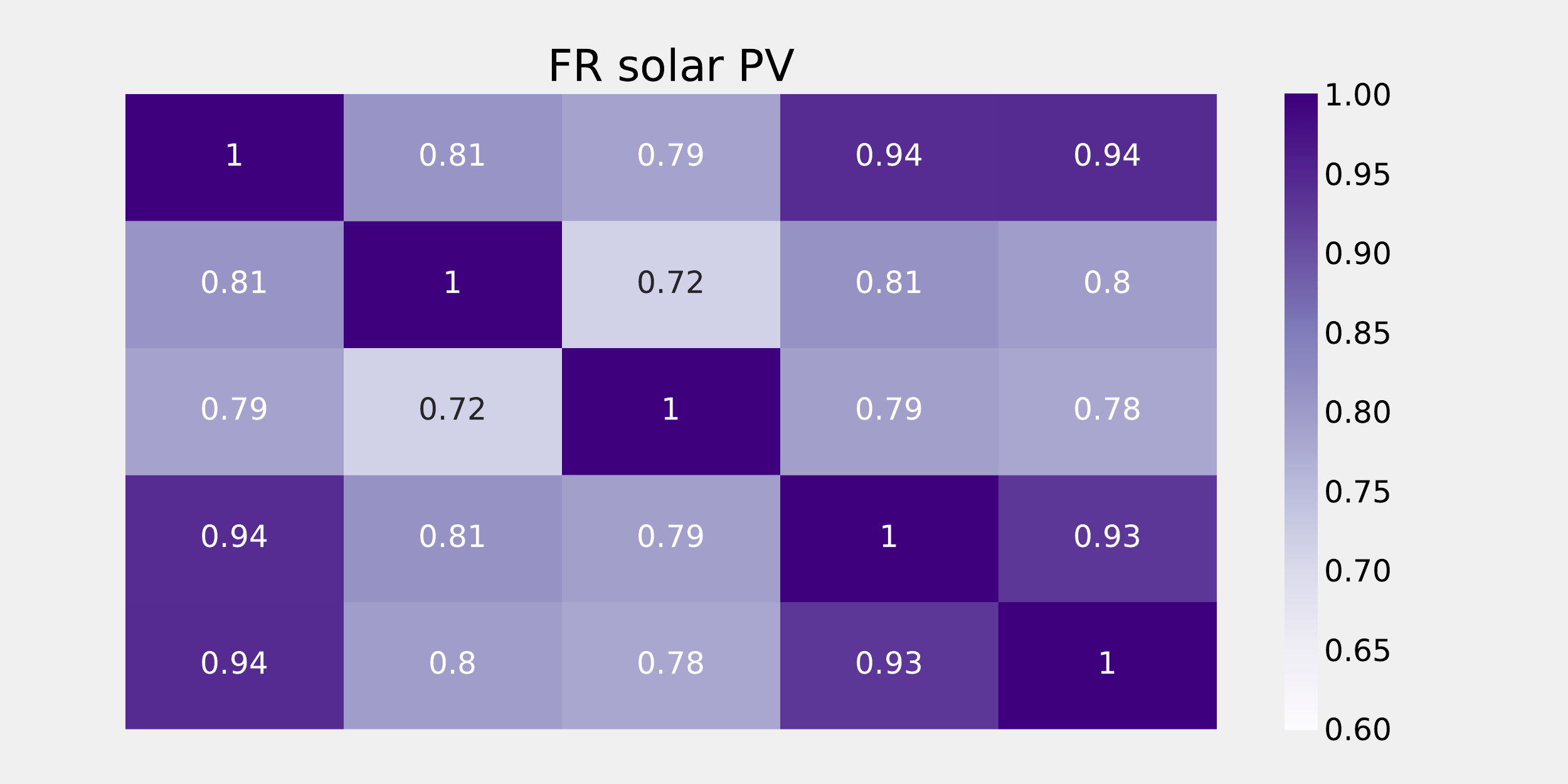}\\\vspace{-0.2cm}
        \includegraphics[width=0.49\textwidth]{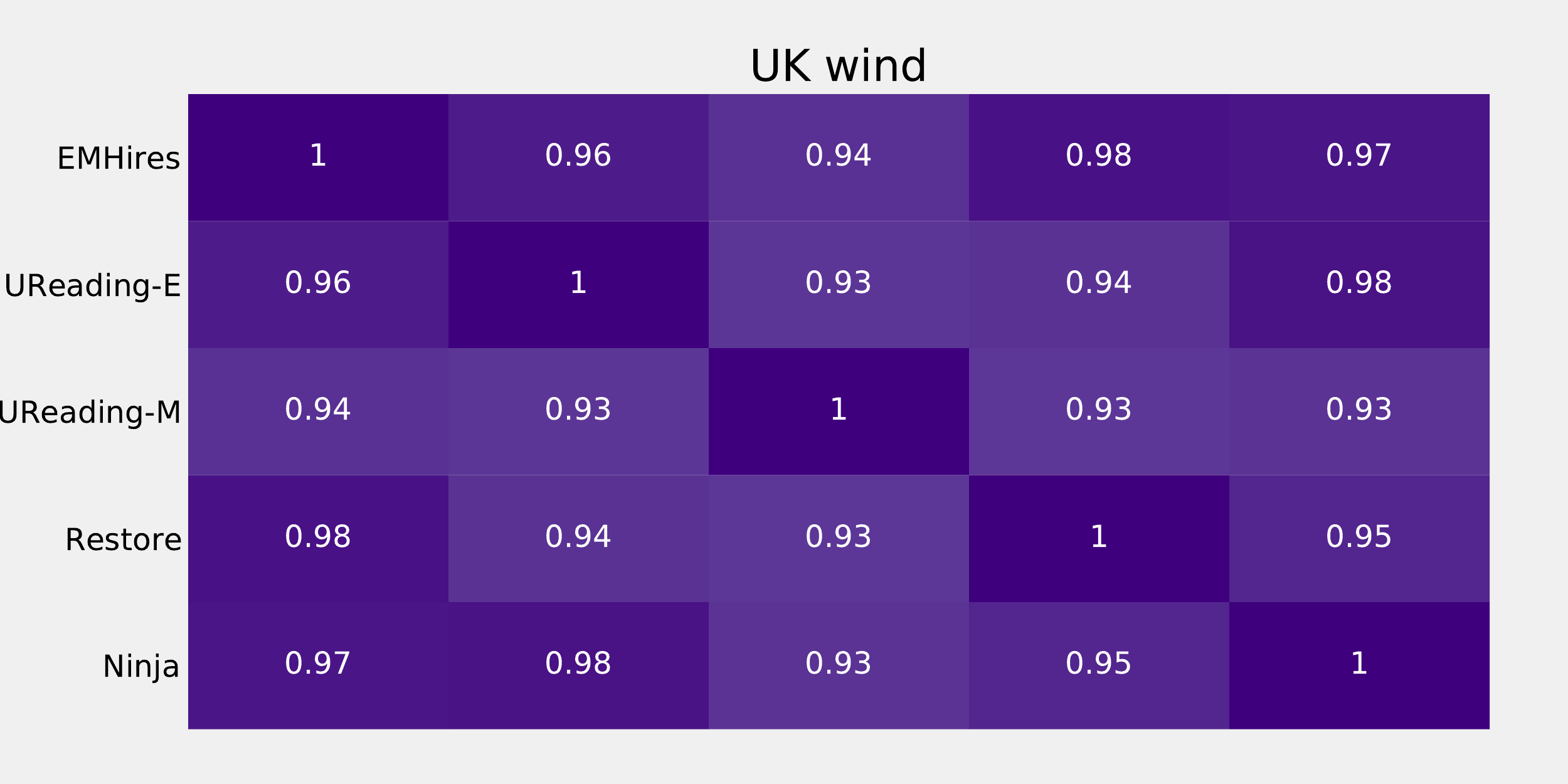}\hspace{-0.2cm}
    \includegraphics[width=0.49\textwidth]{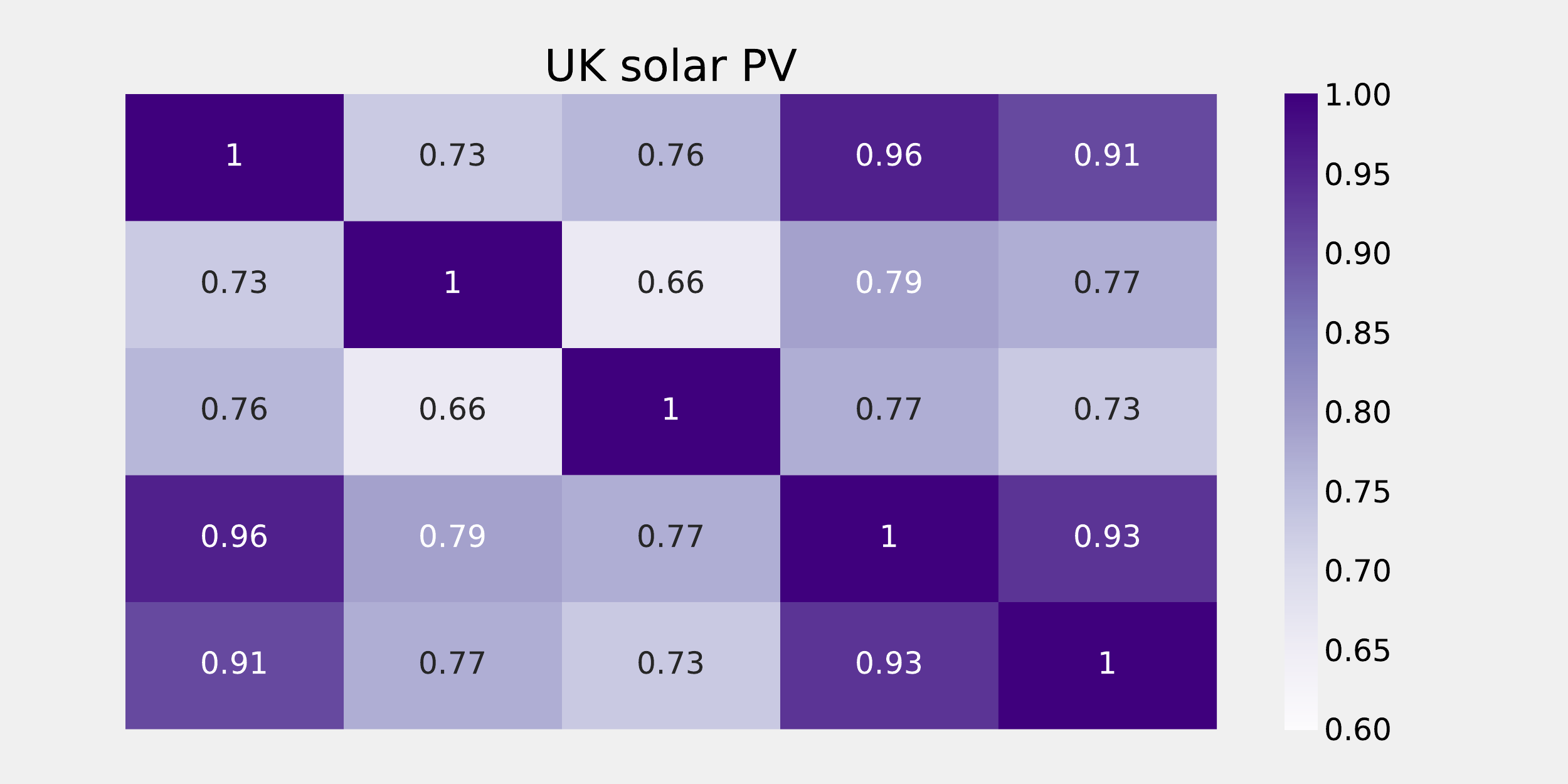}\\\vspace{-0.2cm}
        \includegraphics[width=0.49\textwidth]{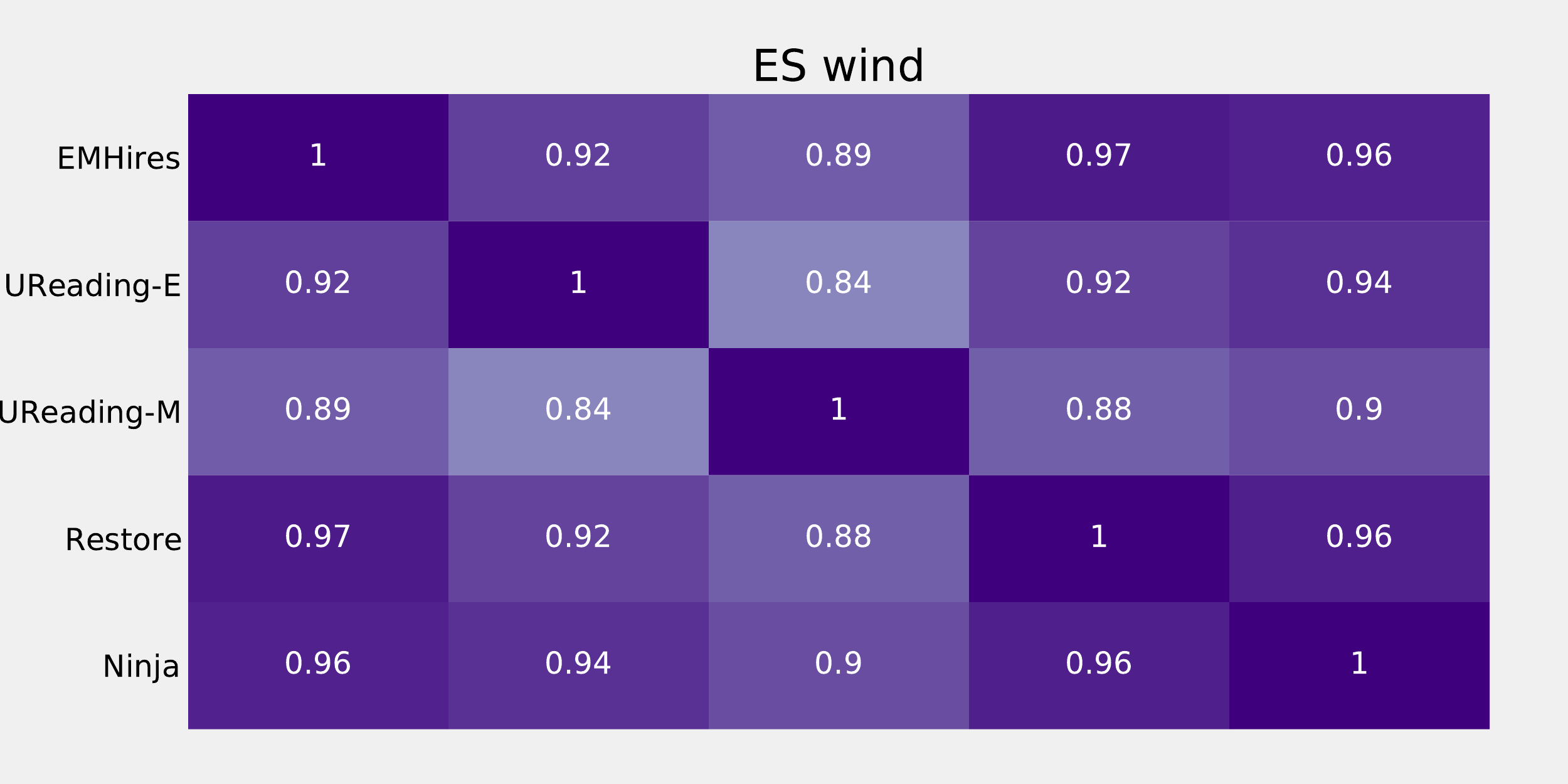}\hspace{-0.2cm}
    \includegraphics[width=0.49\textwidth]{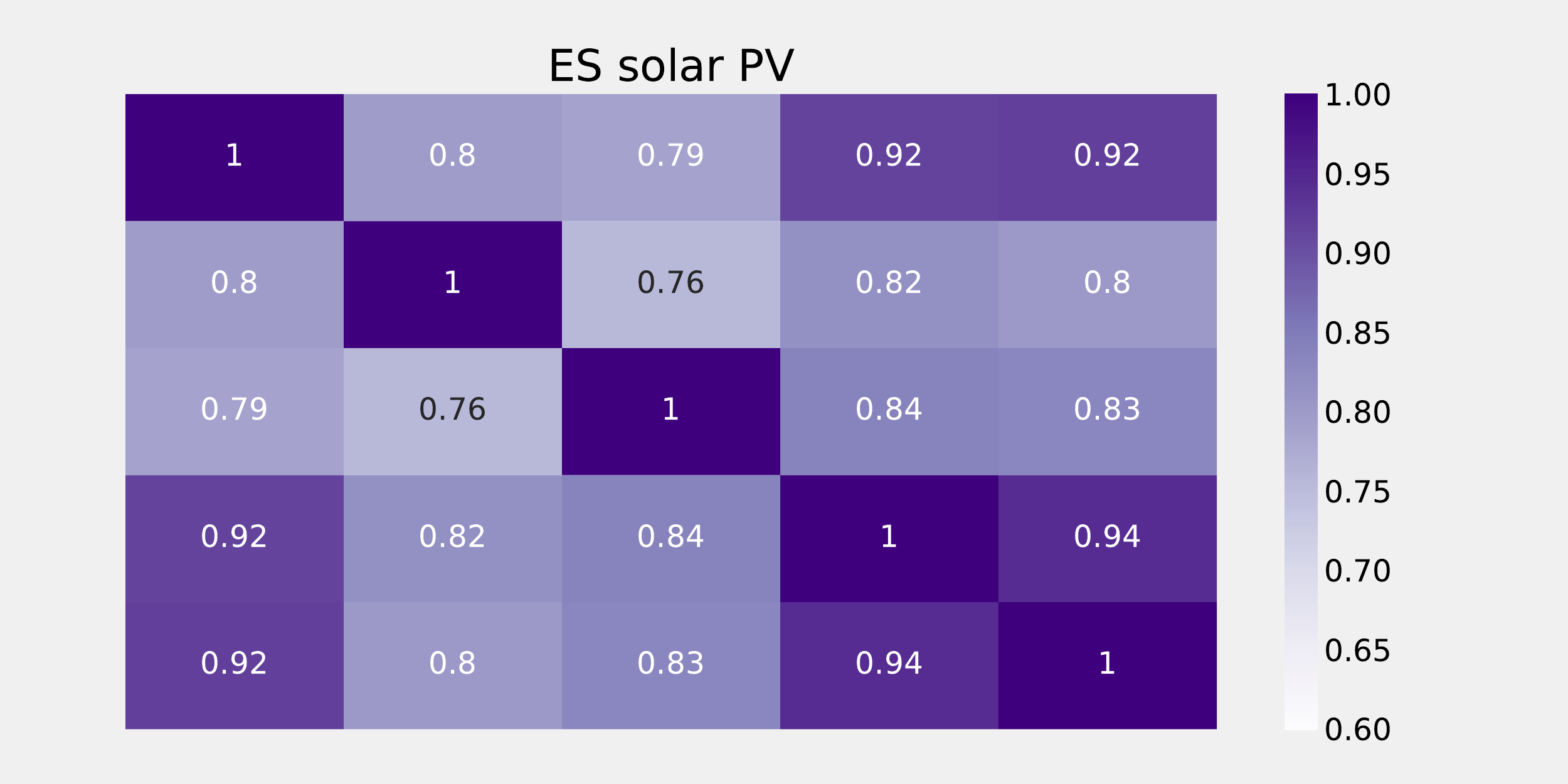}\\\vspace{-0.2cm}
     \hspace{-0.2cm}\includegraphics[width=0.49\textwidth]{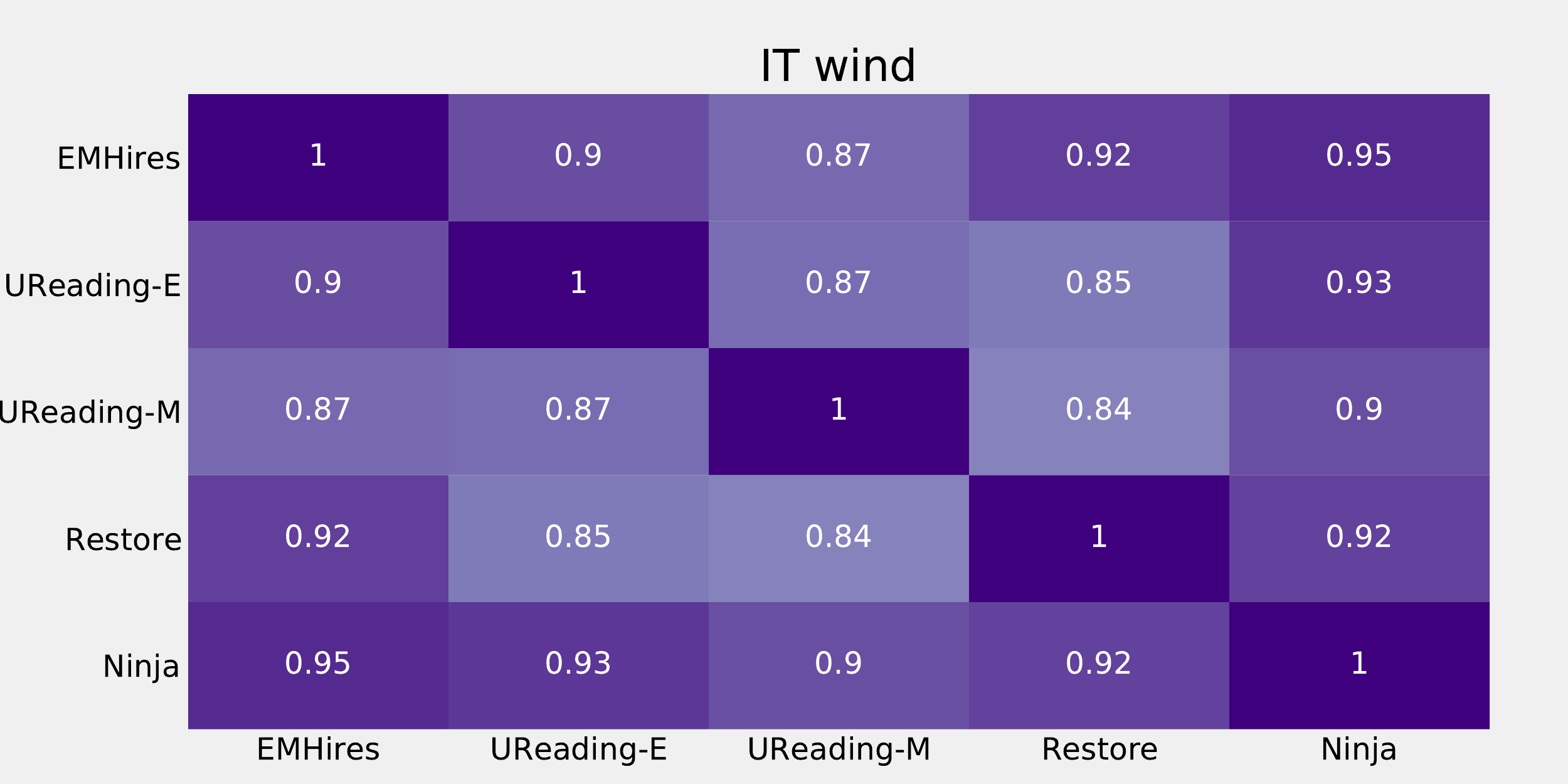}\hspace{-0.2cm}
    \includegraphics[width=0.49\textwidth]{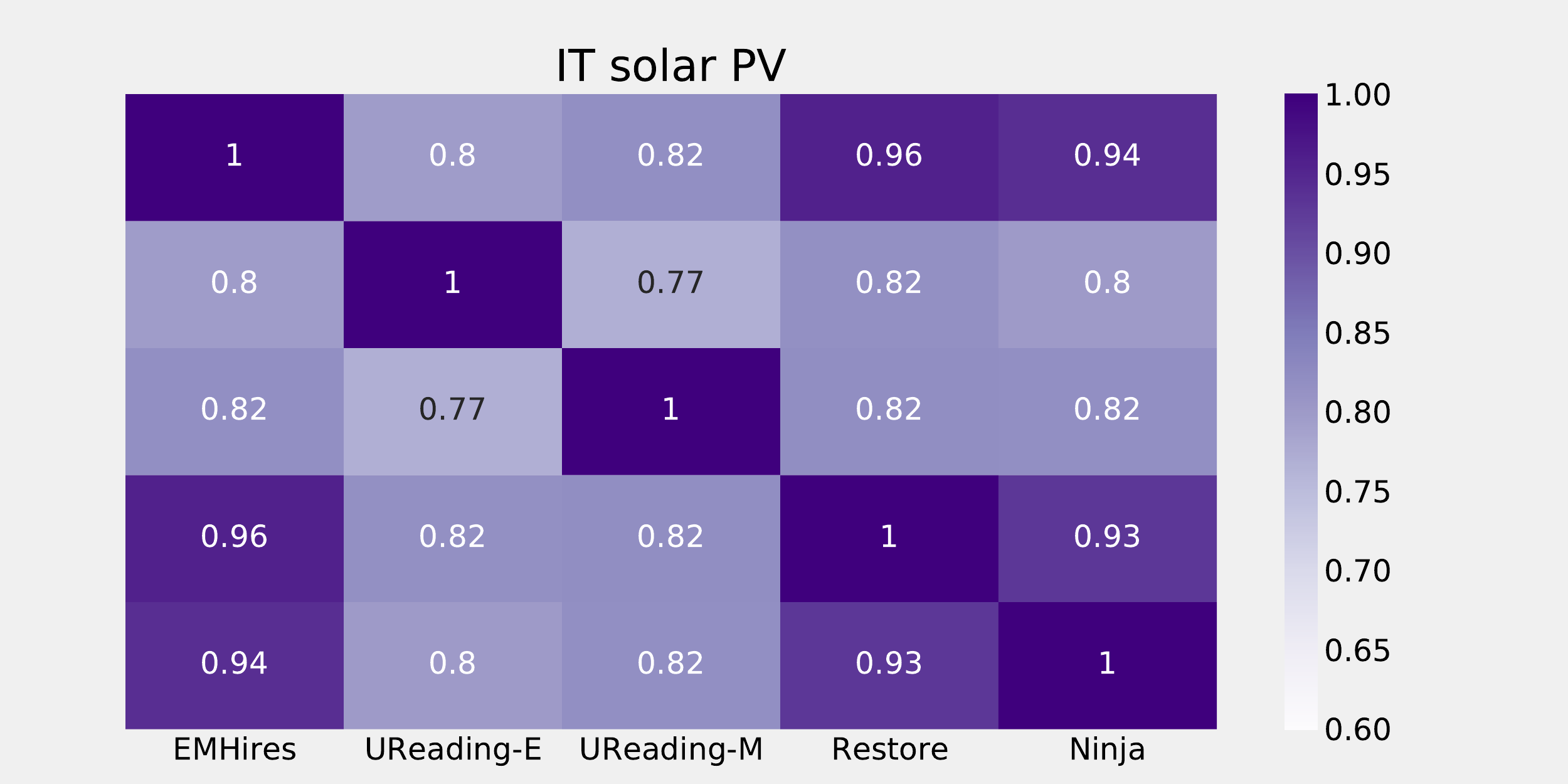}\hspace{-0.2cm}
    \caption{Correlation of locally normalised three-hourly values from 2003-2012. Wind time series were normalised by computing the average and standard deviation over a moving window of 30~days. For solar PV, this was done for each hour of the day separately and a window size of 20 days.}
    \label{fig:correlation}
\end{figure}
 The Pearson correlation coefficient measures linear correlation between two variables.
In the context of renewable generation, the Pearson correlation coefficient is defined as

\begin{align}
    \rho_{g^\alpha,g^\beta} &= \frac{\text{cov}\left(g^\alpha,g^\beta\right)}{\sigma_{g^\alpha}\sigma_{g^\beta}}.
\end{align}

The Pearson correlation coefficient is commonly used to study whether different renewable generation sources or renewable generation sources at different locations are connected via transmission complement each other \cite{giebel2001benefits,jurasz2018impact,francois2016complementarity}.
Here, we are interested in the correlations of the fast (in the order of hours) variations of the wind and solar PV time series. The slow variations caused by large-scale meteorological phenomena and the seasonal cycle of the sun should be similar in all datasets. Therefore, we apply a local normalisation as described in \citet{schafer2010local}: Wind time series are normalised by computing the average and standard deviation over a moving window of 30~days. Solar PV time series are normalised separately for each hour of the day and with a window size of 20~days. This approach was used to reduce the dependency of the time series on the diurnal cycle in case of solar PV and large-scale synoptic conditions for wind.
Figure \ref{fig:correlation} shows pairwise correlation coefficients over the ten years from 2003-2012 for exemplary countries. In general, correlations are very high between datasets despite emphasising the differences by using the moving window approach, except for solar Data from UReading-M, which has comparably low correlations in many cases. However, generally high correlation factors indicate that aggregated temporal patterns are well captured by every dataset. These patterns are important to capture infrastructure requirements that couples different time steps together, such as flexible backup power plant infrastructure or energy storage.

\subsection{Low Generation Events}
Low generation events are times, when resources for both wind and solar PV are not available. In Germany, these events are referred to as "Dunkelflaute", which can be translated as dark doldrums. Low-wind-power events for Germany were studied by Ohlendorf and Schill \cite{ohlendorf2020frequency}. They found that long low-wind-power events are rare. An average wind capacity factor below 10\% for around five consecutive days occurs on yearly basis and for a period of around eight days every ten years. We characterise these events by the aggregated generation from wind and solar PV of all countries together. First, normalised time series are multiplied with installed capacities per country and aggregated

\begin{align}
    G_t = \sum_{n,s} G_{n,s} g_{n,s,t}.
\end{align}

\begin{table}[!h]
\centering
\begin{tabular}{|l|l|l||l|l|l||l|l|l|}
\hline
ctry & wind & solar & ctry & wind & solar & ctry & wind & solar \\\hline
AT  & 2.1  & 0.8   & DE  & 44.9 & 38.2  & PL  & 3.8  & 0.0   \\
BE  & 2.0  & 3.1   & GR  & 2.0  & 2.6   & PT  & 4.9  & 0.4   \\
BG  & 0.7  & 1.0   & HU  & 0.3  & 0.0   & RO  & 3.0  & 1.3   \\
CR  & 0.3  & 0.0   & IE  & 2.3  & 0.0   & SK  & 0.0  & 0.6   \\
CY  & 0.2  & 0.0   & IT  & 8.7  & 18.4  & SI  & 0.0  & 0.3   \\
CZ  & 0.3  & 2.1   & LV  & 0.0  & 0.0   & ES  & 23.0 & 4.8   \\
DK  & 4.8  & 0.6   & LT  & 0.3  & 0.1   & SE  & 5.4  & 0.1   \\
EE  & 0.3  & 0.0   & LU  & 0.1  & 0.1   & UK  & 12.4 & 5.2   \\
FI  & 0.6  & 0.0   & MT  & 0    & 0.1   &     &      &       \\
FR  & 9.3  & 5.6   & NL  & 2.8  & 1.1   &     &      &   \\\hline   
\end{tabular}
\caption{\label{table:capacities}Capacities per country in [GW] \cite{european2015wind,european2015solar} used to produce Fig. \ref{fig:low_gen_events} and Fig. \ref{fig:ramp_rates}.}
\end{table}

Generation capacities by country from wind and solar PV for 2014 are given in Table \ref{table:capacities}.

The resulting time series $\{G_1,G_2,...G_{|t|}\}$ are then split into disjoint subsets $\{G_i,...,G_j\}$
with $G_k < \alpha \sum_{n,s} G_{n,s} \forall k \in [i,...,j]$ and $G_{i-1} >= \alpha \sum_{n,s} G_{n,s} \text{ or } i = 0$ and $G_{j+1} >=\alpha \sum_{n,s} G_{n,s} \text{ or } j = |t|$.
The number of events is then given as the cardinality of the set of subsets $\left|\{\{G_i,...,G_j\},...\}\right|$ and the average length by the average number of elements of this set.
$\alpha$ is the threshold, below which an event is considered to be a low generation event.

\begin{figure}[!htp]
    \centering
    \centering
    \includegraphics[width=0.7\textwidth]{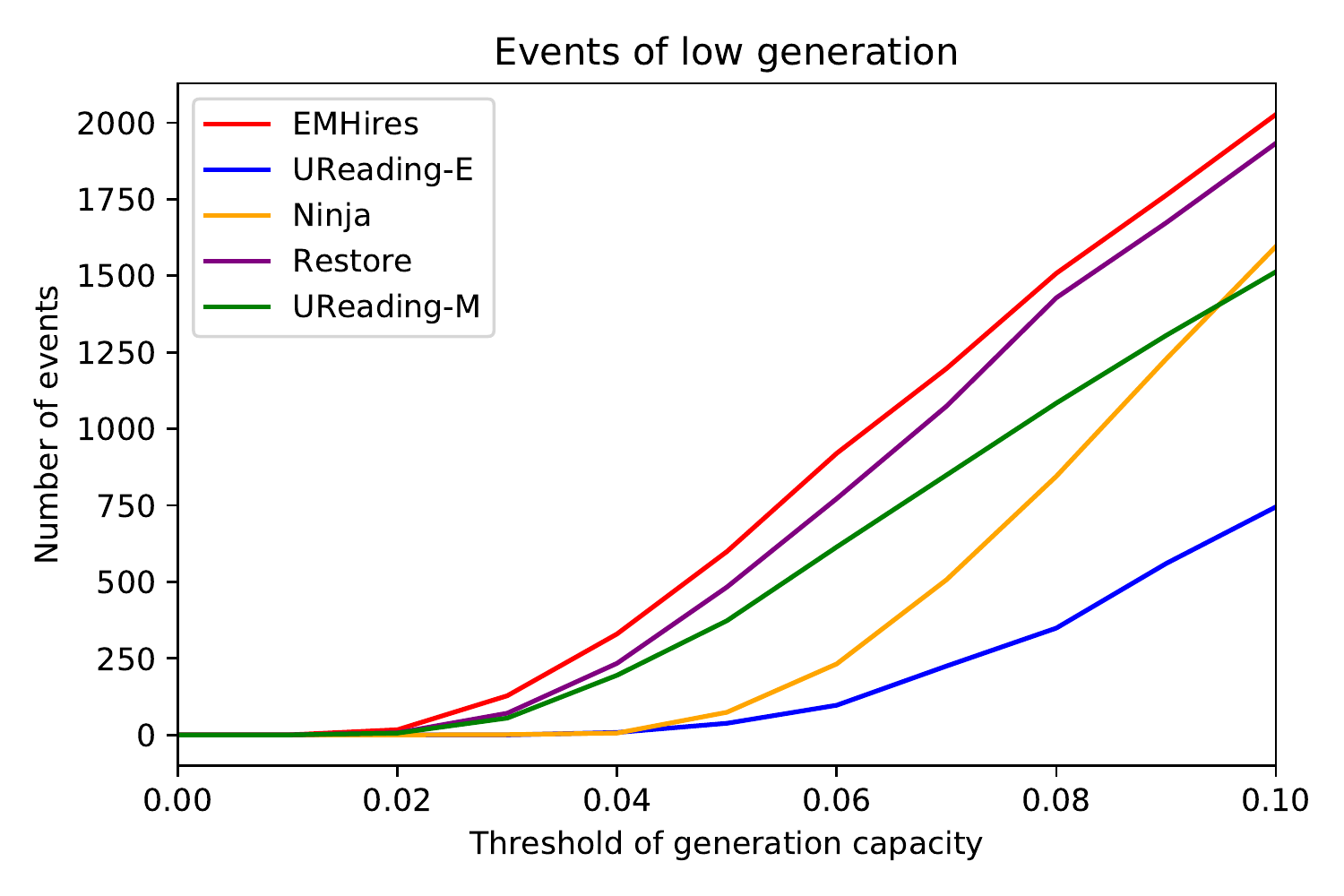}\\
    \includegraphics[width=0.7\textwidth]{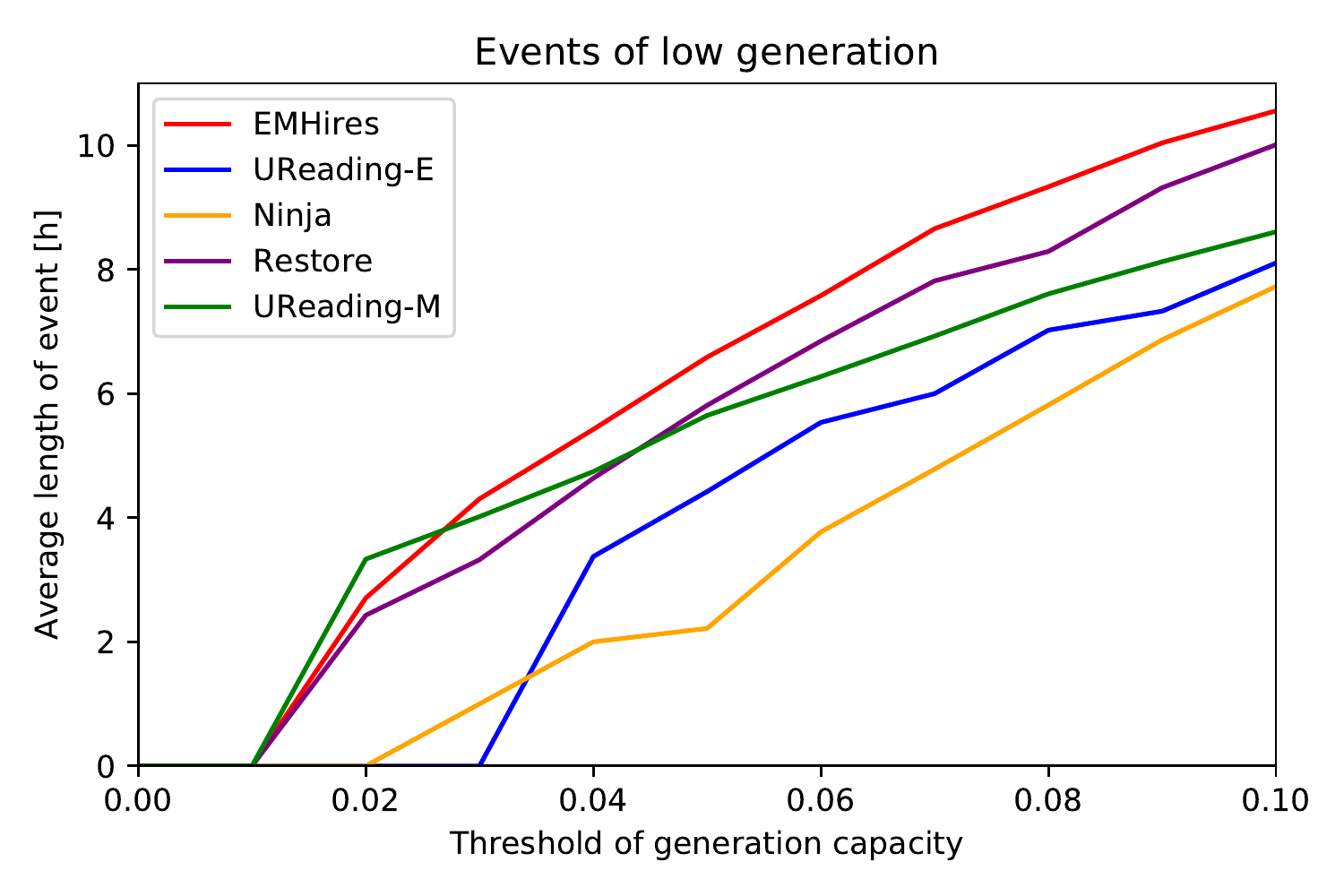}
    \caption{Number of EU28-wide low generation events (top) and their average length (bottom). EMHires shows the highest number of events and length. It should be noted that UReading-E provides only three-hourly data, which makes the direct comparison difficult. Besides, the other ERA5-based datasets show similar slopes for growing thresholds. }
    \label{fig:low_gen_events}
\end{figure}
Figure \ref{fig:low_gen_events} shows the statistics on low generation events for the datasets investigated. Average observed lengths are a few hours and the numbers seems to grow relatively linearly to the growing threshold until a 10\% threshold of total renewable generation. While all datasets show a similar shape, the overall number of events differs quite a lot. It should be noted that for the UReading-E dataset the number of events is influenced by its three-hourly resolution. It is also interesting that the Ninja-dataset demonstrates a stronger growth rate, while most datasets show a similar growth of the number of events with increasing threshold.  Really extreme low generation events are very scarce in Ninja, however the number grows rapidly with the growing threshold overtaking the UReading-M dataset for $\alpha = 0.1$.

\subsection{Ramp Rates}

Another measure of importance for a power system are ramp rates. Ramp rates describe changes in power output that can impact power system operation disproportionately. Batteries are considered to be important components that could deal with ramp rates of renewable generators \cite{frate2019ramp,lappalainen2020estimation}. Ramp rates are often considered in energy resource assessment studies \cite{zhang2018quantitative,widen2011correlations}.
Ramp rates of a power plant are calculated by taking the difference of the output over passed time:

\begin{align}
    RR_{t,\delta_t} = g_{t+\delta_t}-g_{t}.
\end{align}

\begin{figure}[!htp]
    \centering
    \centering
    \includegraphics[width=0.7\textwidth]{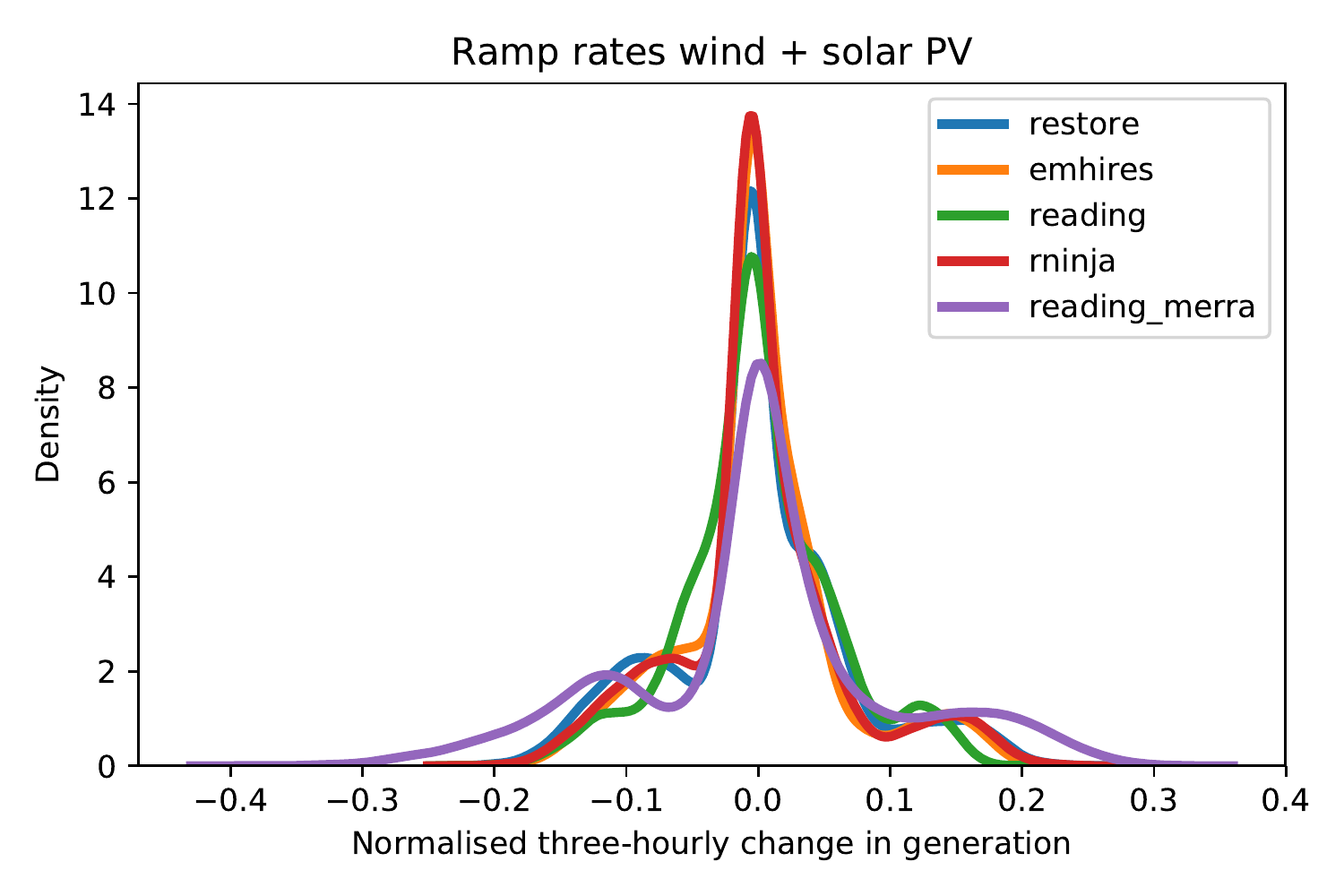}\\
    \includegraphics[width=0.7\textwidth]{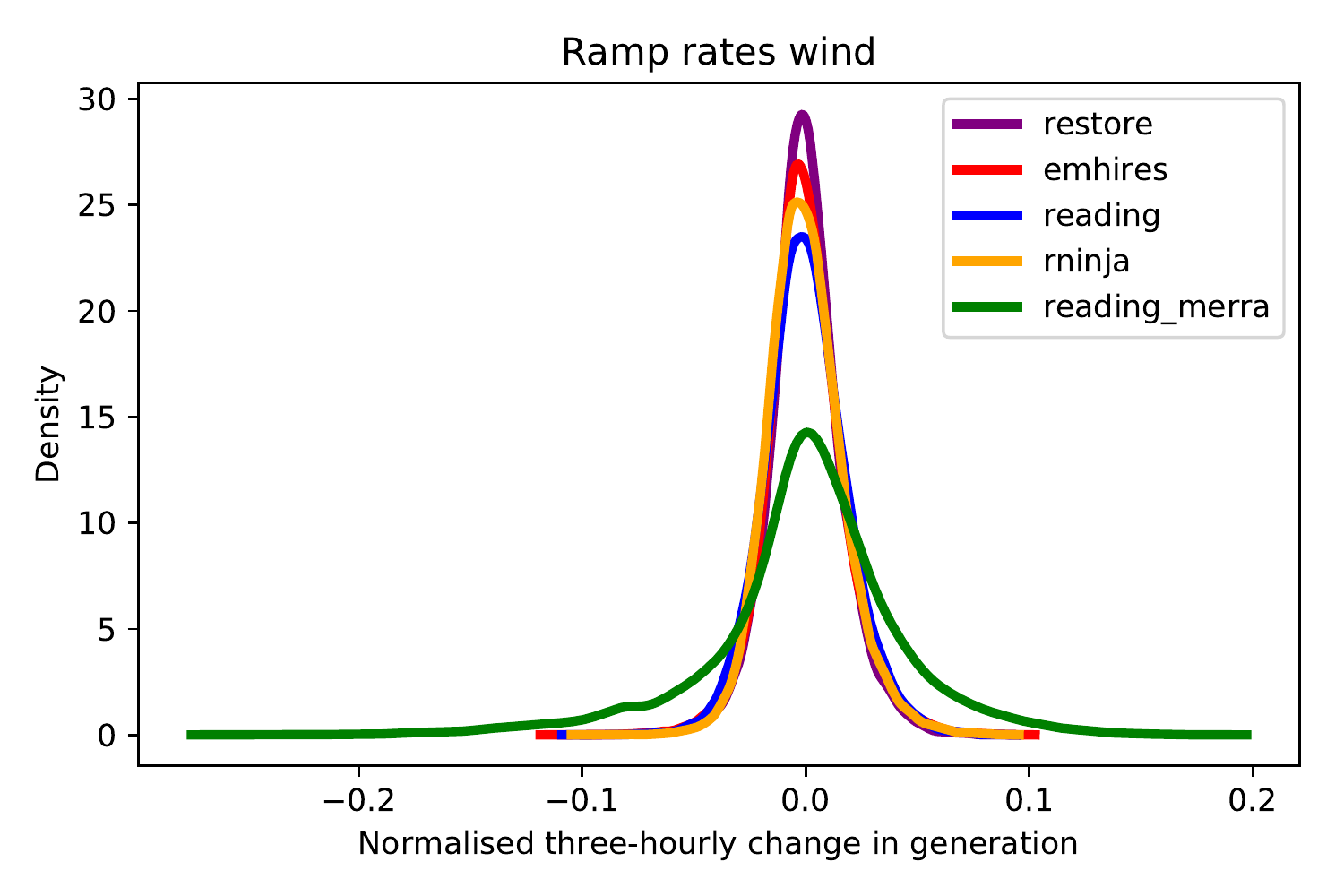}\\
    \includegraphics[width=0.7\textwidth]{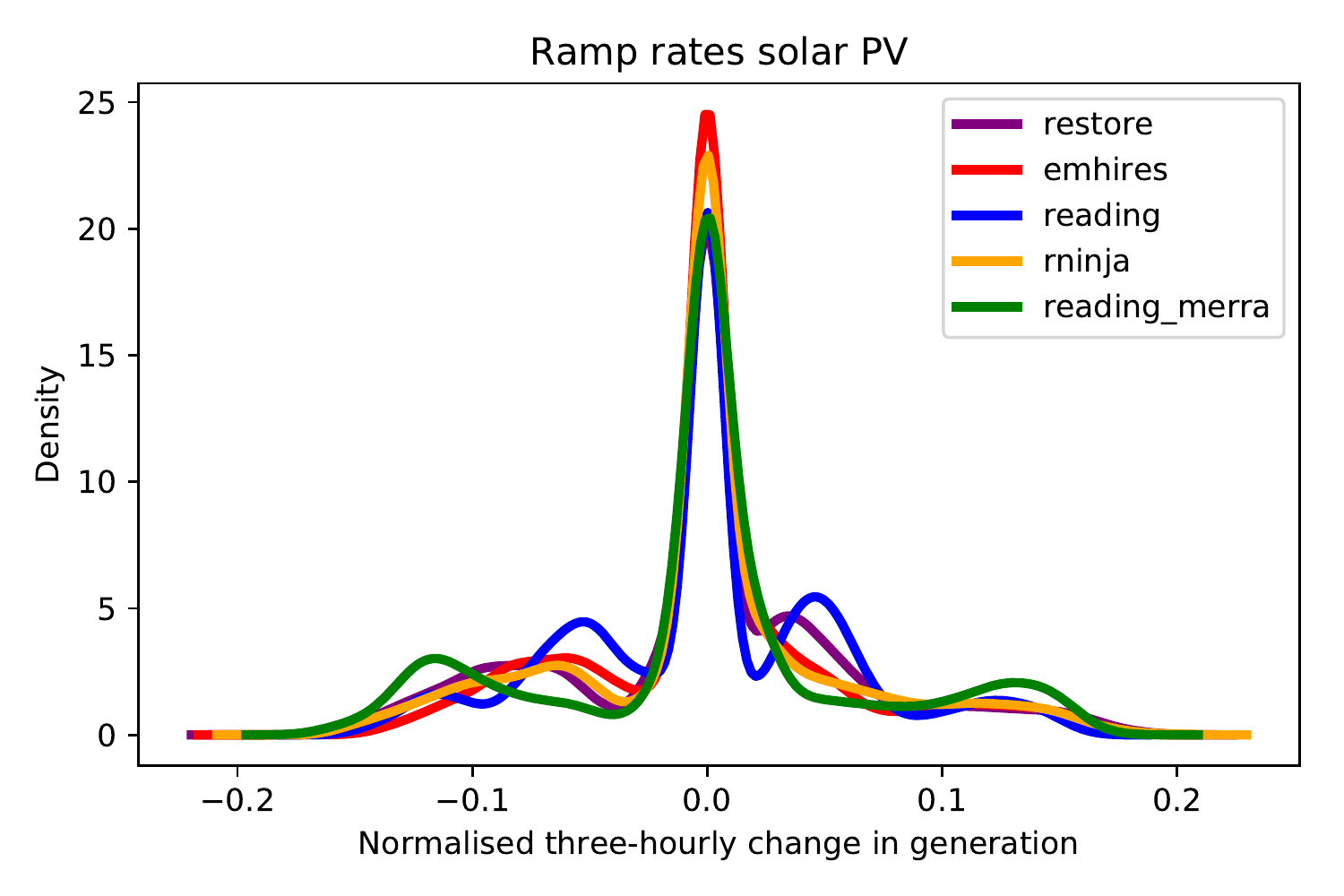}
    \caption{EU28-wide three-hourly ramp rates relative to installed renewable capacities. The small-shoulder peaks for solar PV are likely caused by the deterministic diurnal pattern of the sun. Note that the URead-M data show considerably higher ramp rates for wind than the other datasets. From the system point of view, high ramp rates lead to additional need for balancing.}
    \label{fig:ramp_rates}
\end{figure}

Figure \ref{fig:ramp_rates} shows three-hourly europe-wide ramp rates of the different datasets for wind and PV generation. While the distribution for wind is monotonous and symmetric, the solar PV ramp rate distributions show minor peaks at around $\pm 5$\% of generation capacities. These are likely caused by the deterministic diurnal pattern of PV generation resulting from the rotation of the Earth.
UReading-M ramp rates differ significantly from ramp rates calculated based on other datasets if wind and solar PV are considered separately. However, if wind and solar PV are added together, these effects cancel out and the distribution becomes very similar, with the exception of the relatively prominent second peaks at values $> \pm 0.1$.
\subsection{Optimised Generation Mixes}
\label{ss:optmixed}
Generation data are commonly used to study highly renewable power systems. These include a variety of technologies, such as conventional and renewable generation technologies, storage, transmission technologies, etc. The purpose of power system optimisation is to find the cost-optimal combination of these technologies and their operation under given constraints, for instance carbon dioxide reduction targets. Moreover, models often do not portray the electricity sector alone, but also take heating and transportation into account. This is mostly driven by the eminent need to decarbonize these sectors and use the flexibility they offer to integrate renewables via coupling of the sectors \cite{schaber2013managing,brown2018synergies}.

For renewable generators, the capacity factor of a generator is a crucial variable. One can compute levelized cost of electricity neglecting system measures to balance variable renewable generation as the ratio of cost over generation ($\propto$ capacity factor) \cite{lai2017levelized,talavera2015levelised,bosch2019global}. 
To check the influence of the different datasets on a cost-optimal power system, we present the results of a simplified optimisation in this section.
We optimise the mix of wind, solar PV and open cycle gas turbines (OCGT) as flexible backup power to cover the demand of the EU-28+CH+NO countries. The choice of gas was made because gas is often considered to be a bridge technology towards clean energy that buys time to foster the energy transition \cite{brown2009natural,gonzalez2018review}. 

The optimisation objective reads

\begin{align} \label{eq:minimisation}
 \min_{g,G,F} & \left(\sum_{n,s} c_{n,s} \cdot G_{s} + \sum_{n,s,t} o_{s} \cdot g_{n,s,t} \right) \text{.}
\end{align}
It consists of capital costs $c_{n,s}$ for installed capacity $G_{n,s}$ of a carrier (wind/solar PV) $s$ at node $n$ and marginal costs of generation $o_{s}$ for energy generation $g_{n,s,t}$ of carrier $s$ at node $n$ and time $t$.
It is furthermore subject to the following constraints.

\begin{align} \label{eq:powerbalance}
\sum_s g_{n,s,t} - d_{n,t} = 0 \quad \forall \quad n \text{,} t \text{.}
\end{align}
Demand in space and time needs to be met by dispatched generation from the various generating technologies
and

\begin{align} \label{eq:dispatchconstraint}
0 \cdot G_{n,s} \leq g_{n,s,t} \leq {g}^+_{n,s,t} \cdot G_{n,s} \quad \forall \quad n \text{,} t \text{.},
\end{align}
dispatched generation is limited by generation capacity times a weather-dependent availability ${g}^+_{n,s,t}$ for renewables. This availability is the hourly renewable capacity factor studied in the beginning of the Results section.

\begin{table*}[!h]
\begin{center}
\begin{tabular}{ ||l||r|r|r|| }

    \hline Technology & Capital Cost & Marginal Cost & Emissions \\
                      &  [USD/MW/a] & [USD/MWh] & CO$_2$ [ton/MWh] \\ 
    \hline
    \hline OCGT & 47,235 & 58.385 & 0.635\\
    \hline Wind & 136,428 & 0.005 & 0\\
    \hline Solar PV & 76,486 & 0.003 & 0 \\
    \hline
    
\end{tabular}
\caption{Annualised cost assumptions for generation and storage technologies derived from different sources \cite{carlsson2014etri,schroeder2013current}.}
\label{tab:costassumptions}
\end{center}
\end{table*}

Cost assumptions for generation capacities $c_{n,s}$ and marginal costs $o_{s}$ as well as emission assumptions are given in Table \ref{tab:costassumptions}.

\begin{figure}[!h]
    \centering
    \includegraphics[width=0.9\textwidth]{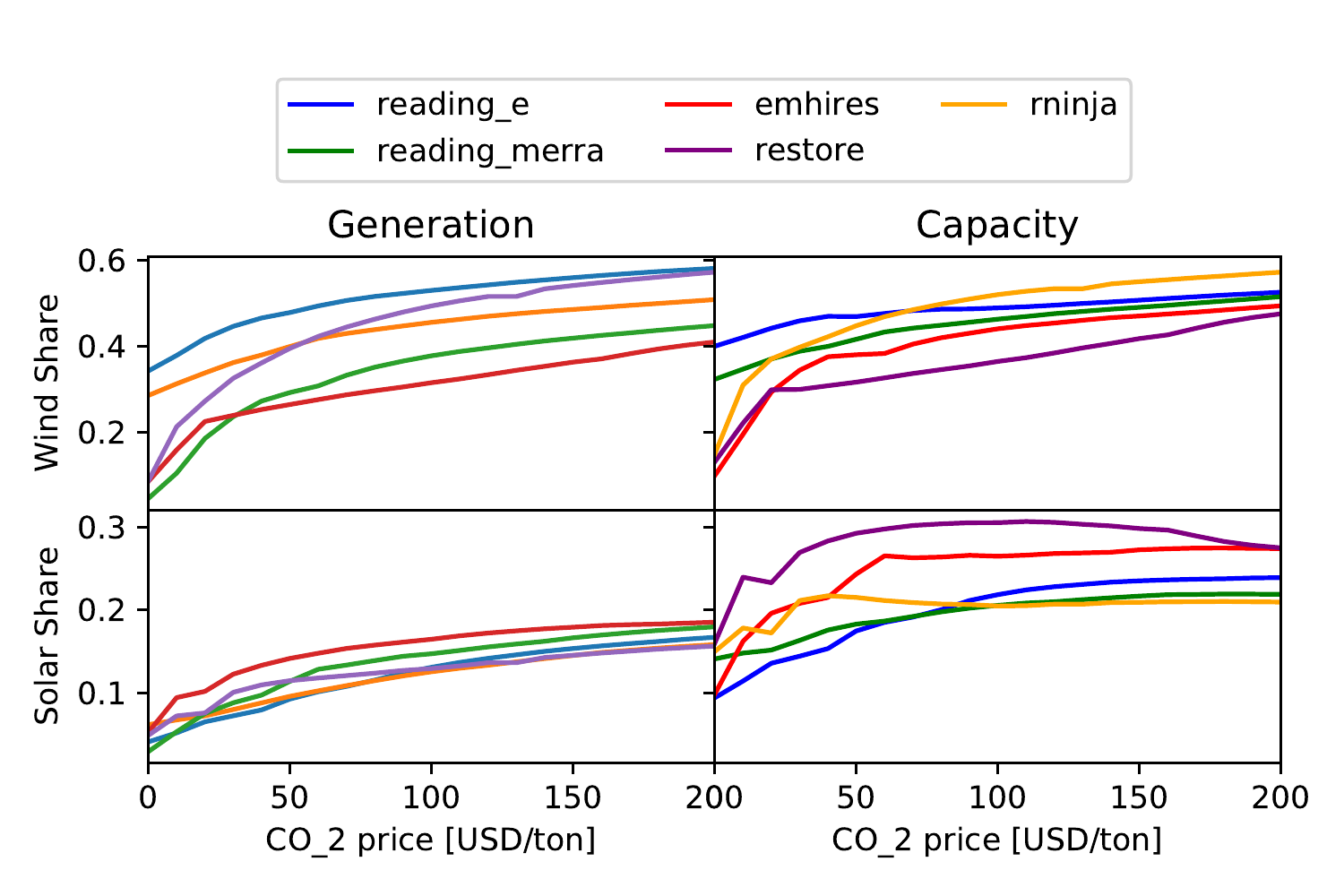}
    \caption{\label{fig:shares}Shares of installed renewable capacities and generation. Capacity curves are less smooth (i.e. monotoneous) than generation curves. Besides, there is a shift in capacity between wind and solar for datasets Restore and Ninja at low CO$_2$ prices indicating a flat optimisation maximum with respect to capacity. The remaining shares comprise gas power plants. }
\end{figure}

Figure \ref{fig:shares} shows shares of the generation and capacities for renewables in dependency of the price for emitting carbon dioxide. This price is multiplied with emissions per generation unit and added to the marginal cost of generation. A growing CO$_2$ price supports the cost-competitiveness of renewables and increases their shares

\begin{figure}[!h]
    \centering
    \includegraphics[width=0.7\textwidth]{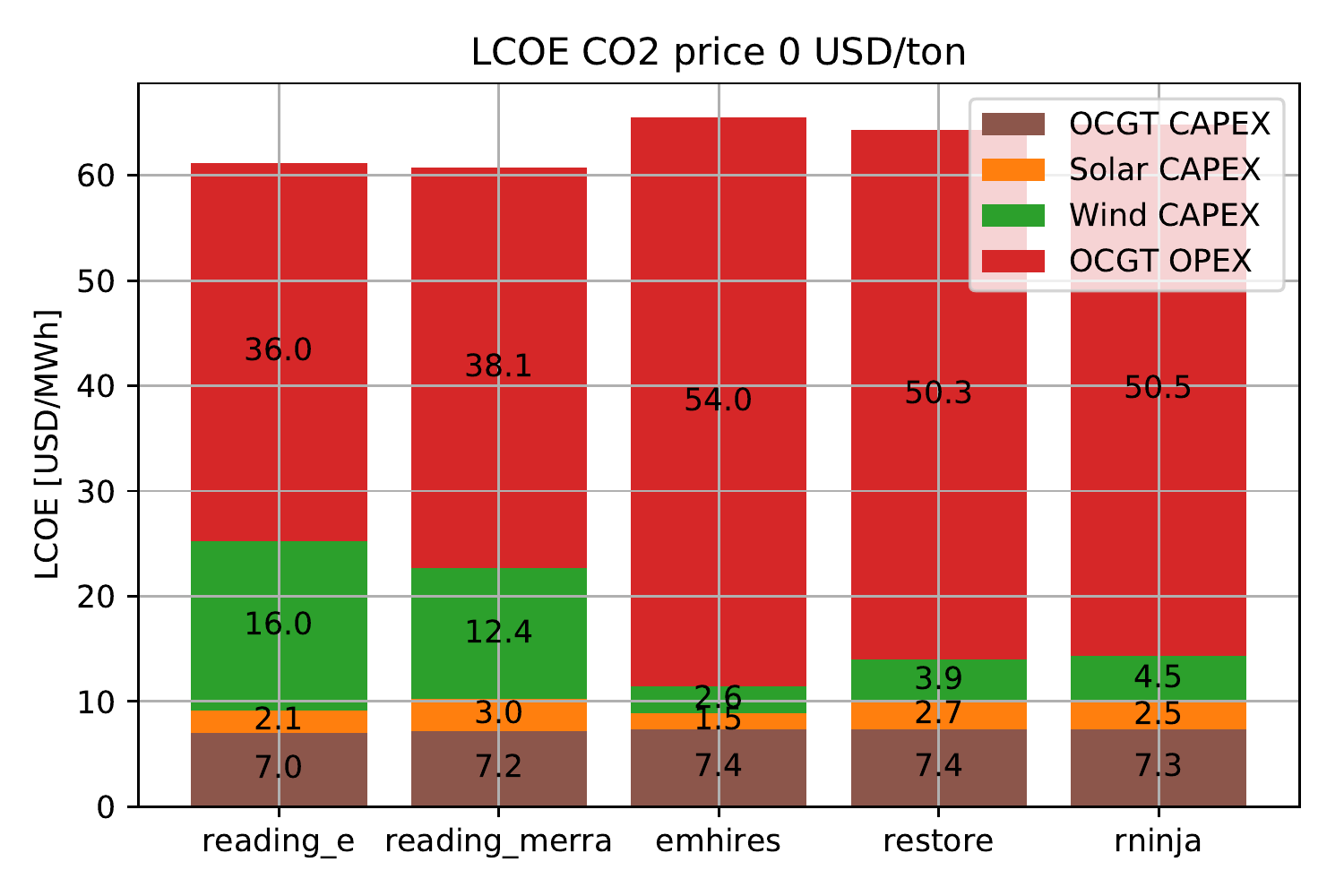}\\
    \includegraphics[width=0.7\textwidth]{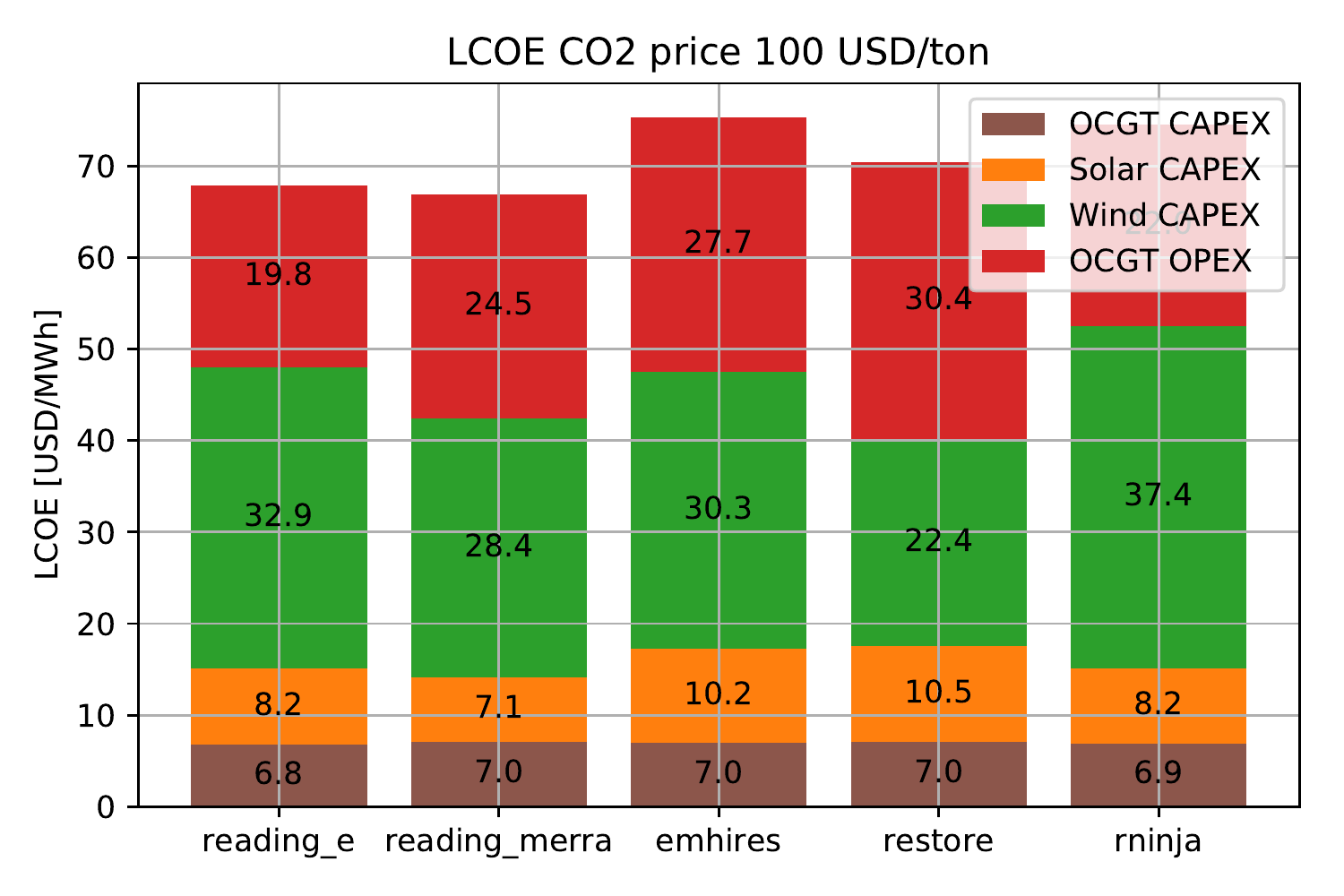}
    \caption{\label{fig:lcoe}Levelized cost of electricity at CO$_2$ prices of 0 USD/ton (top) and 100 USD/ton (bottom) consisting of capital expenditures (CAPEX) and operational expenditures (OPEX), which are neglible for renewables. Note that overall differences are not significantly larger at a high CO$_2$ price. Costs are calculated without the imposed carbon emission price. At a low CO$_2$ price, some datasets (URead-E and URead-M) already render wind generation relatively cost-competitive, while the other datasets do not.}
\end{figure}

\begin{figure}[!htp]
    \centering
    \includegraphics[width=\textwidth]{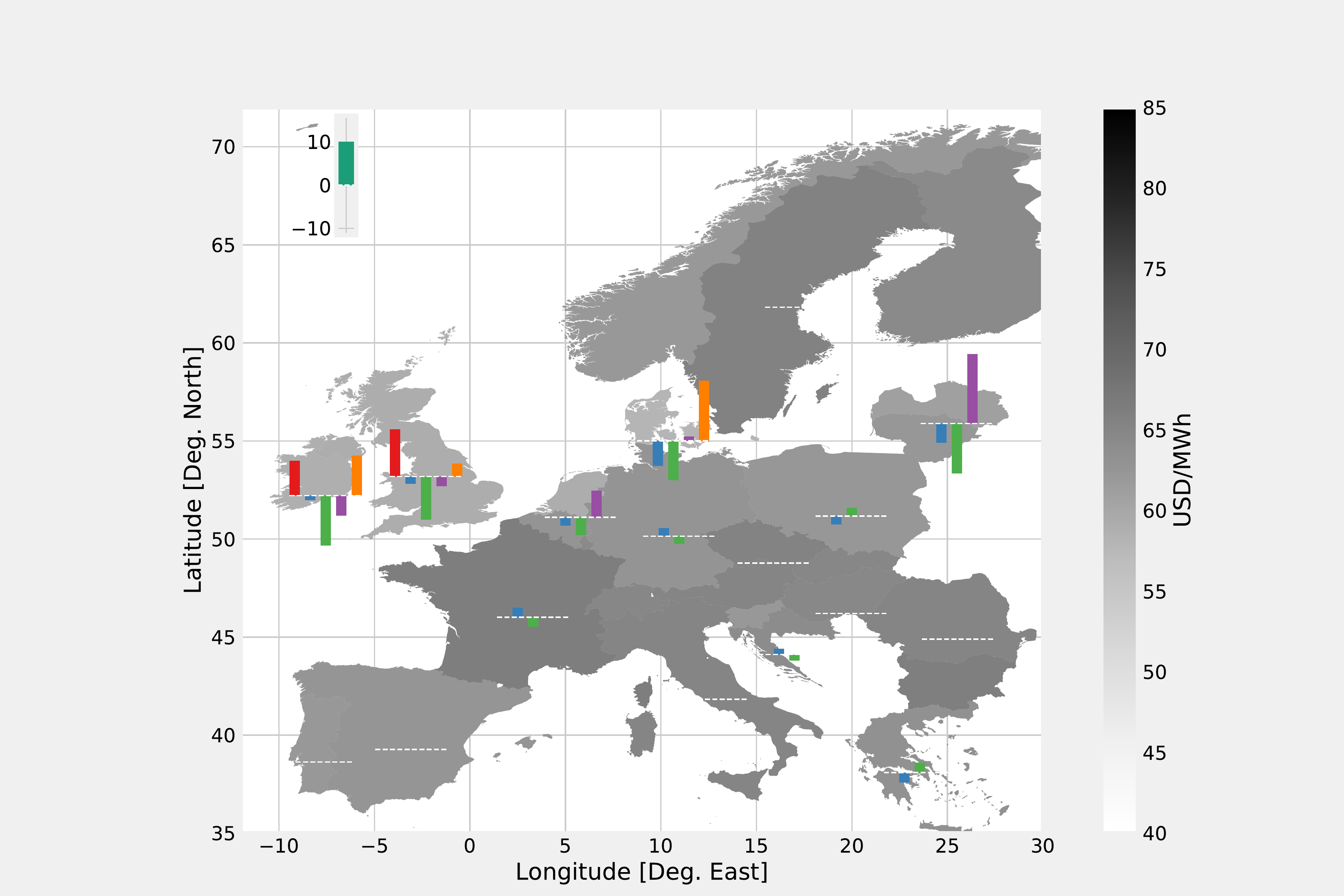}\\
    \includegraphics[width=\textwidth]{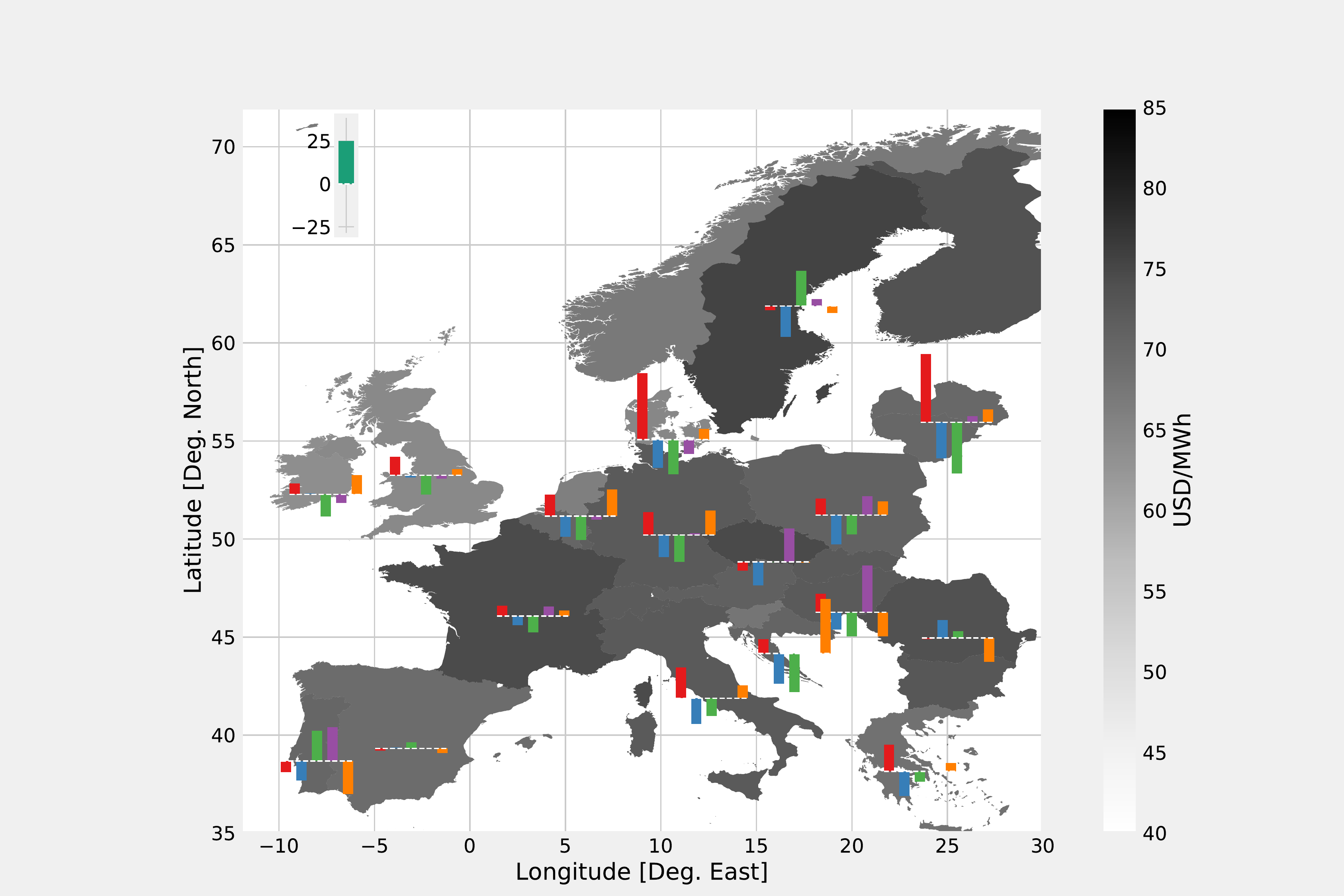}
    \caption{Country-wise LCOE in dependency of the meteorological dataset used for comparison in relation to the ensemble mean. Colors are: EMHires (red), URead-E (blue), URead-M (green), Restore (purple), Ninja (orange) and reported (yellow). These LCOE reflects the cost of autark energy supply per country from renewables plus gas as a backup transition technology. LCOE strongly depends on the datasets at high LCOE prices. Datasets that provide relatively low capacity factors, like EMHires in case of Denmark, lead to a significantly higher LCOE of more than 25\%.}
    \label{fig:lcoe_maps}
\end{figure}

Figure~\ref{fig:lcoe} shows levelized cost of electricity and their composition for CO$_2$ prices of 0 USD/ton and 100 USD/ton without the cost-contribution from emitting carbon dioxide.
Significant differences can be observed. Based on EMHires, Restore or Ninja datasets, renewables are not cost-competitive at a vanishing carbon price, where they contribute only 4-7\%. For the two UReading datasets, the story is different: even at a very low CO$_2$ price, wind contributes  more than 10\% to LCOE indicating relatively large shares of installed capacities. At 100 USD per ton of carbon dioxide emissions, renewables gain significant ground with respect to their cost-competitiveness. Now, for all datasets, wind contributes around 30\% to LCOE and solar PV  around 10\%.

Figure~\ref{fig:lcoe_maps} shows levelized cost of electricity on the country level for the different datasets and carbon emission prices of 0 and 100 USD/ton. At 0 USD/ton, most countries rely entirely on gas, because renewables are not cost-competitive. However, for some countries, such as UK, Ireland, or Denmark, where renewables successfully compete with gas, differences between the datasets are significant. Using datasets such as URead-E and Uread-M that provide relatively high capacity factors, significantly lowers LCOE, by about 10\%. The difference in LCOE for countries that entirely rely on gas is based on differences in the demand patterns, especially between winter and summer.
The effect of the difference in LCOE between datasets is more remarkable at the CO$_2$ price of 100 USD/ton. Cost differences are drastical in some countries such as Denmark, where the results  of different datasets for LCOE can differ by more than 50\%.
The country-resolved plot shows that differences between countries can be significant, if different meteorological datasets for power system optimisation are considered. In case the whole continent is considered, they level out to a large degree.
\section{Discussion}
Data on generation from renewable sources is commonly used in energy system models to study their behaviour, transition pathways or create policy advice. Different data potentially leads to conflicting policy advice and misallocation of large amounts of money.

The integration of renewable energy is a challenging task due to their weather dependency. To cope with non-dispatchable renewable energy sources, various concepts can be applied, such as optimizing the mix of different generation technologies, energy storage \cite{zafirakis2013modeling,zhao2015review}, transmission, demand-side management \cite{nebel2020comparison,kies2016demand} or sector-coupling \cite{schaber2013managing,brown2018synergies}. 
When studying multiple datasets that are commonly used as input into large-scale energy system models, we observed significant differences between the models. As the differences are also observed in datasets based on the same meteorological database, they can likely be attributed to different assumptions in the conversion process to country-aggregated time series. \\

Low generation events are defined as periods of continent-wide low generation from both, wind and solar PV. As these occur continent-wide, a transmission grid does not help to tackle these. Instead, storage is a viable solution. With an average length of a few hours, daily and synoptic storage options seem to be a suitable choice to cope with these low generation events. Commonly, lithium-ion batteries are proposed as a daily storage \cite{jannesar2018optimal} and hydrogen storage or pumped hydro storage (PHS) is a viable storage for the synoptic scale \cite{kroniger2014hydrogen,brown2018synergies}.
Energy system models suffer from uncertainties not only in meteorological data, but also in assumptions made.
Various approaches have been studied to tackle the problem of uncertainty in energy system models
\cite{nacken2019integrated,neumann2019near,schyska2020sensitivity,pedersen2020modeling,hilbers2020importance}.

Besides uncertainty from the generation data source itself and the assumptions made in creating it, additional uncertainty arises from what period to choose. Renewable generation resources tend to vary on decadal scales and models predict that climate change has also profound effects on renewable generation:
Schlott et al. \cite{schlott2018impact} investigated the effects of climate change on a future European energy system and found an increasing competitiveness of solar PV due to changing correlation patterns. Wohland et al. \cite{wohland2017more} found an increasing need of backup energy due to climate change, Weber et al. \cite{weber2018impact} concluded the same and found besides the increasing need for backup energy an increasing need for storage due to climate change. Kozarcanin et al. \cite{kozarcanin201921st} studied various metrics such as variability of renewable generation or short term dispatchable capacity under climate change but concluded that there is no discernible effect on these measures. Bloomfield et al. \cite{bloomfield2020quantifying} saw significant uncertainty in power system design due to climate change and pledged for better understanding of this climate uncertainty.
Besides uncertainty arising from meteorological data, other uncertainties affect power systems, such as uncertainty in cost assumptions as well as technological developments \cite{schyska2020regional,monforti2017comparing}.
The relevance of these effects should be compared to the relevance of uncertainty arising from meteorological data. 

Considering LCOE that were studied in subsection \ref{ss:optmixed} it is interesting to note that differences in the results did not appear to get larger as renewable shares grew. This seems contradictory to the naive estimate that the choice of the renewable generation database should increase the dependency on the choice of meteorological data.
However, in a fully detailed energy system model, complex interdependencies might exist that render smaller or larger effect of different datasets.
One should also note that capacity factors are not the sole factor in determining cost-efficiency of renewables, because non-dispatchable renewables might start cannibalising themselves at high penetration levels reducing their market value \cite{hirth2013market,kies2019market}. Brown and Reichenberg \cite{brown2020decreasing} have argued that this is the result of policy. A possible way to deal with reductions in market value are system-friendly renewable generator designs \cite{hirth2016system,tafarte2020interaction} that aim at producing more at times of higher prices, for instance PV modules with different azimuths/tilts \cite{chattopadhyay2017impact}.

\subsection{Critical Appraisal}
A number of systematic differences increase the difficulty to perform the analysis of the scenarios compared in this paper: datasets are given with different temporal resolution. While one-hourly data are considered sufficient to model renewable country-spanning energy systems with sufficient robustness \cite{brown2018response}, for three-hourly values this is less evident. However, due to the fact that computational limitations and linear optimisation problems are in the complexity class P-complete, we had to find a sweet spot between model details, temporal and spatial resolution \cite{raventos2020evaluation,siala2019impact}  and computational feasibility. At last, we decided to focus on onshore wind. Offshore wind energy was treated differently in the different datasets. However, this effect is rather small, as the installed offshore capacities contribute only a small percentage of the overall wind capacities as of 2014. Nevertheless, its shares are rising quickly, due to technical advances, cost reductions and limited onshore wind potentials \cite{bosch2017temporally,czisch2006szenarien,permien2019socio,mckenna2013determination}, although installation potentials vary significantly in the literature and some researchers suggests far higher numbers \cite{ryberg2019future}. 

\section{Summary and Conclusions}
This paper, compares several distinct datasets, which provide renewable generation time series for both sources, wind and solar PV, on the country level. Different measures are used, such as ramp rates, correlation coefficients, annual capacity factors and optimized mixes of generation, from wind/solar/gas.
From the presented results, the following conclusions can be drawn:
\begin{itemize}
    \item Differences between model statistics are significant, even if they are based on the same meteorological database. These differences are likely due to different assumptions about the conversion from weather to energy. There is a significant need for more research on the effects and interactions of choices made in the modelling chain weather-to-energy at every step, to achieve better understanding.
    \item These significant differences have severe consequences for the optimisation and further studies of power systems. Differences in capacity factors directly affect both, CAPEX and OPEX, of renewable energy generation, transport and storage technologies. Hence their optimized shares in power system expansion models are quite sensible to these uncertainties. 
    \item Emphasis in renewable energy systems research must be put on the use of adequate generation data and discussion of its properties, before we can offer reliable research results and robust policy advice.
\end{itemize}
Future research must focus on the study of these critical issues using a reliable power system model, which allows to capture complex dependencies between both, spatial and temporal effects as well as different types of technologies.
The methods proposed by Schyska and Kies \cite{schyska2020sensitivity}, Nacken et al. \cite{nacken2019integrated} and Neumann and Brown \cite{neumann2019near} shall be considered to improve the study of the effects of the differences in the input weather data on the resulting output of the large-scale energy system models.

\section*{Acknowledgements}
Research is funded by the Federal Ministry of Economic Affairs and Energy (BMWi) under grant nr. FKZ03EI1028A (EnergiesysAI) and the Federal Ministry of Research and Education (BMBF) under grant nr. FKZ 03EK3055C (CoNDyNet II).
HS acknowledges the Judah Eisenberg Professor Laureatus of the Fachbereich Physik and the Walter Greiner Gesellschaft.

\section*{Data Availability}
The datasets analysed in this work are available in a harmonised form under https://github.com/alexfias/compare\_met\_data/.

\bibliographystyle{elsarticle-num-names}
\bibliography{sample.bib}







\end{document}